\documentclass[twocolumn,prd,nofootinbib,aps,floats,floatfix,amsmath,amssymb,secnumarabic]{revtex4} %
\usepackage[final]{graphicx}
\usepackage{amsmath}
\usepackage{amsfonts}
\usepackage{amssymb}
\usepackage{latexsym}
\usepackage{graphicx}
\usepackage[english]{babel}
\usepackage{multirow}
\usepackage{float}
\usepackage{url}
\usepackage{hyperref}
\usepackage{slashed}
\usepackage{xcolor} 

\def\sfrac#1#2{{\textstyle{#1\over #2}}}
\newcommand{\be}{\begin{equation}}
\newcommand{\ee}{\end{equation}}
\newcommand{\ba}{\begin{array}}
\newcommand{\ea}{\end{array}}
\newcommand{\bea}{\begin{eqnarray}}
\newcommand{\eea}{\end{eqnarray}}
\newcommand{\sss}{\scriptscriptstyle}

\def\sfrac#1#2{{\textstyle{#1\over #2}}}

\begin{document}

\title{Consistency of dark matter interpretations of the 3.5 keV X-ray line}
\author{James M.\ Cline}
\affiliation{Department of Physics, McGill University,
3600 Rue University, Montr\'eal, Qu\'ebec, Canada H3A 2T8}
\author{Andrew R.~Frey}
\affiliation{Department of Physics and Winnipeg Institute for Theoretical 
Physics, University of Winnipeg, Winnipeg, Manitoba, Canada R3B 2E9}

\begin{abstract}

Tentative evidence of a 3.5 keV X-ray line has been found in  the
stacked spectra of galaxy clusters, individual clusters, the Andromeda
galaxy and the galactic center, leading to speculation that it
could be due to decays of metastable dark matter such as sterile
neutrinos.  However searches for the line in other systems such as
dwarf satellites of the Milky Way have given negative or ambiguous
results. We reanalyze both the positive and negative searches from the
point of view that the line is due to inelastic scattering of dark
matter to an excited state that subsequently decays---the mechanism of
excited dark matter (XDM).  Unlike the metastable dark matter scenario,
XDM gives a stronger signal in systems with higher
velocity dispersions, such as galaxy clusters.  We show that the
predictions of XDM can be consistent with null searches
from dwarf satellites, while the signal from the closest individual
galaxies can be detectable having a flux consistent with that from
clusters.  We discuss the impact of our new fits to the data for two
specific realizations of XDM.

\end{abstract}
\maketitle

\section{Introduction}  

A surprising new hint of dark matter emerged from analysis of data
from XMM-Newton, in which the spectra of 73 galaxy clusters were
combined, showing $> 3\sigma$ evidence for an X-ray line with energy
3.55 keV \cite{Bulbul:2014sua}.  It was argued that there were no
plausible atomic transitions to account for such a line, but that it
could come from the decay of light dark matter (DM) such as sterile
neutrinos.   Further evidence for the line was found in the spectra of
the Perseus cluster and (less prominently) of the Andromeda galaxy
\cite{Boyarsky:2014jta}. A subsequent search for the line in the
center of the Milky Way  using Chandra data gave negative results 
\cite{Riemer-Sorensen:2014yda} whereas a similar search using 
XMM-Newton data corroborated the line \cite{Boyarsky:2014ska}. Ref.\
\cite{Anderson:2014tza} stacked spectra of 81 and 89 galaxies using
Chandra and XMM-Newton data, respectively, finding no evidence for the
line, while ref.\ \cite{Malyshev:2014xqa} searched for the line in
stacked spectra of nearby dwarf spheroidal galaxies, also with
negative results.  The latter two papers emphasize that there is a
definite contradiction to the decaying dark matter interpretation made
by ref.\  \cite{Bulbul:2014sua}, at the level of $3.3 - 4.6$ $\sigma$
for ref.\ \cite{Malyshev:2014xqa} and $4.4 - 11.8$ $\sigma$ for ref.\
\cite{Anderson:2014tza}.  Previous searches for X-ray
lines are reviewed in ref.\ \cite{Boyarsky:2012rt}.
Most recently (after the first version of this work), 
ref.\ \cite{Urban:2014yda} reported a positive flux using 
Suzaku data from the Perseus
cluster but only upper limits from other nearby clusters.

There is also controversy as to whether an atomic origin for the line
is really excluded.  Ref.\ \cite{Jeltema:2014qfa} argues that
transitions of ionized potassium and chlorine explain the line 
reported by \cite{Bulbul:2014sua,Boyarsky:2014jta}.  Counterarguments
have been given in \cite{Boyarsky:2014paa,Bulbul:2014ala}.  We do not
enter into this debate in the present paper (however see 
ref.\ \cite{Iakubovskyi:2014yxa} for a recent synopsis). Instead we will assume that
the line is due to new physics, namely dark matter scattering rather
than decays.  The kinematical differences between the two processes
can explain why an X-ray line would be seen in some data sets and not
in others.  In particular, if the scattering is inelastic with
a small energy threshold,
one expects the strongest signal to come from  from galaxy
clusters, while that from dwarf galaxies would be highly suppressed,
and line strengths from nondwarf galaxies would be somewhere
between these two extremes.

As a concrete realization of this alternative phenomenology,  we focus
on a class of dark matter models in which inelastic  scattering of two
DM particles to excited states, $\chi \chi \to \chi'\chi'$, is
followed by rapid decays $\chi'\to \chi\gamma$
\cite{Finkbeiner:2007kk,Pospelov:2007xh,Finkbeiner:2014sja, Frandsen:2014lfa}.  
We refer to this as the
excited dark matter (XDM) mechanism.  The DM need not be as light as
7.1 keV, as in the metastable decaying models; it can be heavy,
requiring only the mass splitting between $\chi'$ and $\chi$ to be
3.55 keV. Specific examples of XDM models for addressing the 3.55 keV
X-ray signal were considered in refs.\ \cite{Cline:2014eaa,Cline:2014kaa}.

The papers that searched for the line signal derived limits (or
observed ranges) for the mixing angle $\theta_\nu$ of a Majorana sterile neutrino, 
decaying through its transition magnetic moment to an active neutrino.
The mixing angle is related to the partial width for the decay by
\cite{Boyarsky:2005us}
\bea
	\Gamma_\nu &=& {9\,\alpha\, G_F^2\over 1024\,\pi^4}\,
	\sin^2 2\theta_\nu\,
	m_\nu^5\nonumber\\
	&=& 2.46\times 10^{-28}{\rm s}^{-1}\,
	{\sin^2 2\theta\over 10^{-10}}\,\left(m_\nu\over 7.1\,
	{\rm keV}\right)^5
\eea
Ref.\ \cite{Bulbul:2014sua} finds the best fit for $\sin^2
2\theta\cong 6\pm 3\times 10^{-11}$.  Ref.\ \cite{Boyarsky:2014jta}
obtains a consistent result, with larger errors.  Ref.\ 
\cite{Anderson:2014tza}
finds the $3\sigma$ upper limit $\sin^2 2\theta\ < 2\times 10^{-11}$,
while \cite{Malyshev:2014xqa} obtains $\sin^2 2\theta\ < (2.7-4.8) \times 10^{-11}$,
depending upon different assumptions about the contribution from 
decays of DM in the main halo of the Milky Way.  Ref.\ 
\cite{Riemer-Sorensen:2014yda} finds somewhat weaker
limits $\sin^2 2\theta\ < (5-10) \times 10^{-11}$, depending upon the
energy interval that is modeled.\footnote{\label{rsr} These numbers are from 
fig.\ 4 of the revised version of ref.\ \cite{Riemer-Sorensen:2014yda}
provided to us by the author.  We recompute the limits based upon our
own assumptions about the DM density profile for table \ref{tab1}.} These results are summarized in
table \ref{tab1}.

\begin{table*}[ht]
\begin{tabular}{|l|c|l|c|c|l|l|}
 \hline
$\qquad$ (1) & (2) & $\ (3)\ \nu$ mixing & $\ $(4) fast decay & 
(5) intermediate & 
$\ $(6) slow decay &(7) $v$
disp.\\
 $\quad\quad$ Reference & object &  $\quad\sin^2 2\theta_\nu $ &
${\langle\sigma v\rangle_f\cdot \left(10\, {\rm GeV}\over
m_\chi\right)^2}^{\phantom{|}}$& $\tau\sim 2\times 10^6$y
& ${\langle\sigma v\rangle_s\cdot \left(10\, {\rm GeV}\over
m_\chi\right)^2}^{\phantom{|}}$ &
$\quad\langle \sigma_v\rangle$\\
 &  & $\ (\times 10^{-11})$
&  ($10^{-22}\,{\rm cm}^3{\rm s}^{-1}$)  & or\ $2\times 10^7$y & ($10^{-22}\,{\rm cm}^3{\rm s}^{-1}$) & (km/s)\\
\hline
Bulbul {\it et al.} \cite{Bulbul:2014sua} & clusters & $\quad\ \ 6\pm 3$ &
 $\quad 480\pm 250$  & & $\quad\  1200\pm 600$ & $\ \ \ 975$ \\
\hline
Bulbul {\it et al.} \cite{Bulbul:2014sua} & Perseus & $\quad (26-60)$ &
 $\quad (1400-3400)$ & &  $\ \ (4000-15000)$ & $\ \ 1280$ \\
 \hline
Boyarsky {\it et al.} \cite{Boyarsky:2014jta} & Perseus & 
$\quad (55-100)$ & 
$\ \ (1-2)\times 10^{5}$  & & $\ \ \ (1-5)\times 10^{4}$& $\ \ 1280$\\
\hline
Urban {\it et al.} \cite{Urban:2014yda} & Perseus &  $\quad (20-100)$ & 
$\ \ (2600-4100)$  & & $\ \ \ (1-2)\times 10^{4}$& $\ \ 1280$\\
\hline
Bulbul {\it et al.} \cite{Bulbul:2014sua}
&CCO\footnote{Coma+Centaurus+Ophiuchus clusters}& $\quad(18-28)$ & 
$\quad(1200-2000)$  &   & $\quad(5100-8400)$   & $\ \ \ 926$ \\
\hline
Boyarsky {\it et al.} \cite{Boyarsky:2014jta} & M31 &  $\quad(2-20)$ & 
$\quad\ \left\{\begin{array}{ll}(10-30),& {\rm NFW}\\ 
(30-50),& {\rm Burkert}\end{array}\right.	$ & 
${{\rm NFW}\atop\to}\left\{ {{\rm unchanged}\atop (20-50)}\right\}$ 
& $ \ \quad(370-970)$& $\ \ \ 116$\\
 \hline
Boyarsky {\it et al.} \cite{Boyarsky:2014ska} & MW & $\quad (10-30)$ &
$\quad\left\{\begin{array}{ll}(0.1-0.7),& {\rm NFW}\\
	(50-550),& {\rm Burkert}\end{array}\right.$
 & ${{\rm NFW}\atop\to}\left\{ { (1-8)\atop (16-110)}\right\}$ 
& $\quad\ (400-3000)$ & $\ \ \ 118$ \\ 
\hline
\hline
Riemer-S\o rensen \cite{Riemer-Sorensen:2014yda} & MW & $< (6-20)$ & 
$\ < \left\{\begin{array}{ll}(0.15-1.1),& {\rm NFW}\\ (80-1200),& {\rm
Burkert}\end{array}\right.$
 & ${{\rm NFW}\atop \to}\left\{{ (2-12)\atop (24-170)}\right\}$ & $\
< (200-2000)$& $\ \ \ 118$\\
\hline
Anderson {\it et al.} \cite{Anderson:2014tza} & galaxies & $< (2-5)  $ & $\ <
(270-620)  $ &  &  $\ < (170-420)$ & $\ \ \ 100$\\
\hline
Malyshev {\it et al.} \cite{Malyshev:2014xqa} & dwarfs &  $< (3-5)$ & 
$\ < (0.2-0.3)$ & &  $\ < (0.1-0.2)$ & $\ \ \ \ 10$\\
\hline 
Bulbul {\it et al.} \cite{Bulbul:2014sua} & Virgo & $<(18-23) $ &
 $<(380-670) $  & & $\ <(2.5-4.1)\times 10^4$ & $\ \ \  643$ \\
\hline
Urban {\it et al.} \cite{Urban:2014yda} & Coma &  $<(1.5-1.7)$ & 
$\ \ <(130-200)$  & & $\ < (510-850)$& $\ \ \  913$\\
\hline

\end{tabular}
\caption{Column 3: best-fit values or upper limits on the sterile neutrino
mixing angle, $\sin^2 2\theta$, assuming $\nu_s\to\nu\gamma$ for the
3.55 keV X-ray line. Column 4: corresponding values of the cross
section for excited dark matter models with $\chi\chi\to
\chi'\chi'\to\chi\chi\gamma\gamma$ for the case of prompt 
decay of $\chi'$.  For fast decays of XDM in the Milky Way,
fits are given both to NFW and Burkert profiles.
Column 5: for the case of excited state lifetimes $\tau \sim 2\times
10^6\,$y  or $2\times 10^7\,$y
the MW cross sections change relative to NFW values in column 4 as shown,
while others are unaffected.
Column 6: same as column 4 but for slow decays
(lifetime of order the dynamical time scale).
Column 7: average velocity
dispersion.  Values for Coma, Centaurus clusters from ref.\ 
\cite{Girardi:1995iy}, and 
for Ophiuchus from \cite{2010MNRAS.405.1624M}.}
\label{tab1}
\end{table*}

In this work, we will systematically derive the corresponding
values for the  phase-space-averaged cross section
$\langle\sigma v\rangle$ that plays the role of $\Gamma_\nu$ for the XDM
scenario.  We do using the predicted X-ray fluxes from the two
models:\footnote{There is no factor of two in front of $\langle\sigma
v\rangle$ to account for two photons being produced by the
two decays following $\chi\chi\to\chi'\chi'$, since we assume $\chi$
to be Majorana.  In this case the factor of two is canceled by a
factor of $1/2$ for having identical particles in the initial state.}
\bea
\label{fluxeq1}
	F_\gamma &=& \quad\,\Gamma_\nu
	\left<\int {d^{\,3} x\over 4\pi\,x^2 }\,{\rho\over
	m_\nu}\right>,
	\quad {\rm decays}\\
\label{fluxeq2}
	 &=& \langle\sigma v\rangle_{f} \left<\int{d^{\,3} x\over
4\pi\,x^2 }\,
	{\rho^2\over m_\chi^2}\right>,\quad{\rm XDM}
\eea
Here $\rho$ is the DM mass density, the origin of $\vec x$ is at the 
observer, and $m_\nu = 7.1$ keV for decaying
DM, while $m_\chi$ can be much larger in the XDM model.  
The subscript on $\langle\sigma v\rangle_{f}$ refers to the assumption
that $\chi'$ decays relatively fast, as we will discuss further below.
The large 
angle brackets indicate that an average over different sources is
typically being carried out, be they dwarf galaxies, normal galaxies,
or clusters of galaxies.  By performing this average in the same way
for XDM as it was carried out by the original authors we can convert
their determinations of $\Gamma_\nu$ into corresponding values for 
$\langle\sigma v\rangle$.  However in most cases we can work directly from
the reported fluxes using eq.\ (\ref{fluxeq2}).  
This is the first objective of our work.  
We will then show how the expected DM velocity dependence of 
$\langle\sigma v\rangle$ can make the derived values consistent with an
XDM origin for the observed line.

We identify a small discrepancy in the inferred cross sections (and neutrino
mixing angles) for the Milky Way and Andromeda galaxy (M31), despite similar
velocity dispersions, if we assume a standard NFW profile.  However, if the
Milky Way's DM halo is slightly cored rather than cuspy, the amount of DM
in the field of view is reduced, leading to a larger value of the required 
cross section.  Since a similar change does not affect the line-of-sight
integral (\ref{fluxeq2}) much for M31 due to the greater distance, we find
that slight coring removes the discrepancy.

So far we implicitly assumed that the excited state decays immediately,
but it is also possible that it could be sufficiently 
long-lived that it migrates significantly before decaying.  In the
extreme case where the lifetime is of the same order as the 
dynamical time scale for the object of interest, the excited states become distributed evenly
throughout the halo, and 
the brightness profile of the X-ray line
has the same shape as for decaying DM, although the overall 
predicted rate 
differs from that of purely decaying DM.
In that case the photon flux takes the same form as in (\ref{fluxeq1}),
but with the replacements $m_\nu\to m_\chi$ and $\Gamma_\nu\to \Gamma_{\rm
eff}$, where the effective decay rate given by 
\be
	\Gamma_{\rm eff} = \left<
	{\langle\sigma v\rangle_s\,\int {d^{\,3}
x}\,\rho^2\over
	m_\chi\,\int {d^{\,3} x}\,\rho}\right> = 
{\langle\sigma v\rangle_s \over
	m_\chi} \left<\rho_s{h_2(c)\over h_1(c)}\right>
\label{gamma_eff}
\ee
with the integrals extending out to the virial radius $R_{\rm vir}$
of the halo. $\rho_s,\ r_s$ are the respective scale density and length for the DM distribution, to be defined
below, and $h_{1,2}(c)$ are dimensionless functions of the concentration
parameter $c= R_{\rm vir}/r_s$, given in the appendix.  The subscript
on $\langle\sigma v\rangle_s$ denotes that $\chi'$ is assumed to decay
slowly.  In the following we will carry out our analysis for both extremes
of the excited state lifetime, as well as intermediate cases.  We will
show that lifetimes of order $10^6-10^7\,$y 
provide an alternate resolution to the discrepancy between the galactic 
center and M31 fluxes.

In the remaining sections  \ref{clusters}-\ref{galaxies}
we determine the required values or upper limits for the
cross sections from galaxy clusters, M31, the Milky Way, the Perseus
cluster, dwarf spheroidals, and stacked galaxies respectively.
In section \ref{compat} we show how these can be fit to general
parameters of the XDM class of models.  The implications for 
two specific models are considered in section \ref{models}, followed
by our conclusions.

\section{Galaxy clusters}\label{clusters}
  Ref.\ 
\cite{Bulbul:2014sua} combines spectra of 73 galaxy clusters.
For each cluster the integral $\int d^{\,3}x\, \rho/ x^2$ 
(called $M^{proj}_{DM}/D^2$ in \cite{Bulbul:2014sua}) is determined 
 within a given field of view (FOV),
defined by an extraction radius $R_{ext,i} = \theta_i d_i$ where
$\theta_i$ is the angular size of the observed region and $d_i = D$ is
the distance to the source.  
The relative exposure ${Exp}_i$ for each source is also given.

For each cluster, an NFW profile is assumed,
\be
	\rho_{\rm\sss NFW} = {\rho_{s,i} \over (r/r_{s,i})(1+r/r_{s,i})^2}
\label{NFW}
\ee
The scale radius is taken to be $r_{s,i} = R_{500,i}/c_{500}$ where 
$R_{500,i}$
is the radius of a sphere whose average density 
is $500$ times the critical density of 
the universe, and the concentration parameter is a weakly-varying 
function of the virial mass that is in the range 
$c_{500}\sim 3-4$ for most clusters\cite{vik}.  
For the more distant clusters, $R_{500,i} =
R_{ext,i}$, which is tabulated in ref.\ \cite{Bulbul:2014sua}.
But for the closer ones, $R_{500}$ exceeds the XMM-Newton $700''$
field of view (FOV).  In order to determine the NFW parameters for
all clusters in a consistent way, we assume 
(as did  \cite{Bulbul:2014sua}) a universal value of the concentration
parameter, which implies that all clusters have the same value of
$\rho_s = \delta_c\rho_{\rm
crit}$, where the critical density is given by $\rho_{\rm crit} =
1.25\times 10^{11}M_\odot/$Mpc$^3$ (ignoring small redshift-dependent
corrections) and $\delta_c = (500/3)c^3/h_1(c)$ (see for example ref.\ 
\cite{Urban:2014yda}). We then determine $r_s$ for each cluster by equating the
projected mass tabulated in \cite{Bulbul:2014sua} to 
$M^{proj}_{DM} =  4\pi \rho_s r_s^3 f_1(R_{\rm ext}/r_s)$,
where $f_1$ is given in eqs.\ (\ref{fieqs}).

Once the parameters of $\rho_{\rm\sss NFW}$ are known, it
is straightforward to compute $\int d^{\,3}x\, \rho^2/x^2$ for each
cluster
and average them, weighting by the exposures ${Exp}_i$.  Taking
$\Gamma_\nu = 1.74\times 10^{-28}\,$s$^{-1}$ corresponding to ref.\ 
\cite{Bulbul:2014sua}'s best-fit mixing angle, we obtain from
eqs.\ (\ref{fluxeq1}-\ref{fluxeq2}) the cross section
$\langle\sigma v\rangle_f = 5.5\times 10^{-20}\,$cm$^3\,$s$^{-1}
\times(m_\chi/10{\rm\ GeV})^2$, in the case of promptly decaying
excited states, assuming $c=3$, and $4.1\times 10^{-20}\,$cm$^3\,$s$^{-1}
\times(m_\chi/10{\rm\ GeV})^2$ for $c=4$.  This gives some idea as to
the uncertainties $(\sim 15\%)$ due to the DM halo properties.  In table \ref{tab1}
we combine this in quadrature with the significantly larger $(50\%)$ 
uncertainty from the measured flux. 

For slowly decaying $\chi'$, we equate
$\Gamma_\nu/m_\nu = \Gamma_{\rm eff}/m_\chi$ from (\ref{gamma_eff})
and (\ref{fluxeq1}) to find 
\be
	\langle\sigma v\rangle_s = \Gamma_\nu {m_\chi^2\,h_1\over
	m_\nu\,h_2}
	\rho_s^{-1}
\ee
leading to the higher values
$(0.95-1.4)\times 10^{-20}\,$cm$^3\,$s$^{-1}$ for $c=3-4$.

\section{Andromeda Galaxy}  Ref.\ \cite{Boyarsky:2014jta} attributes a
flux of $F_\gamma = 4.9{+1.6\atop-1.3}\times 10^{-6}$/cm$^2$/s to a
combined FOV of radius 0.22$^\circ(\equiv\!\theta)$ centered on M31.  
We follow \cite{Boyarsky:2014jta} in
adopting the best-fit  NFW profile of ref.\ \cite{Corbelli:2009nc} for
M31, with  $r_s = 23.8\,$kpc and $\rho_s = 4.3\times
10^6\,M_\odot$/kpc$^3$.\footnote{These follow from the concentration
parameter $c=12$ for an overdensity of $\Delta=98$ and $M_\Delta =
1.2\times 10^{12}M_\odot$, using $M_{\sss\Delta} = 
\sfrac{4\pi}{3}R_{\sss\Delta}^3\rho_c\Delta = 4\pi\rho_s
R_{\sss\Delta}^3 h_1(c)$ with $c = R_{\sss\Delta}/r_s$.}\  Taking the
distance to be $d=778\,$kpc, hence $a=\theta d/r_s = 0.13$,
we compute the integral of $\rho^2/x^2$
and find the cross section  $\langle\sigma v\rangle = m_\chi^2 F_\gamma
d^2/(\rho_s^2 r_s^3 f_2(a)) = 2.1\times 10^{-21}\,$cm$^3\,$s$^{-1}
\times(m_\chi/10{\rm\ GeV})^2$ for fast decays. Alternatively, ref.\
\cite{klyp} finds best-fit values of $r_s = 25\,$kpc (also with $c=12$,
and $R_{\rm vir}=300\,$kpc, $M_{\rm vir}= 1.6\times 10^{12}M_\odot$) and
$\rho_s = 5.0\times 10^6\,M_\odot$/kpc$^3$, giving $1.4\times
10^{-21}\,$cm$^3\,$s$^{-1}$.  The uncertainty due to the density profile
is somewhat smaller than that from the line flux for the overall
uncertainty in the cross section.   In table \ref{tab1} we give the range
of values that includes both uncertainties.  

For slowly-decaying DM, we find, using 
\be
	\langle\sigma
	 v\rangle_s = {F_\gamma\, m_\chi^2\, d^2\,  
	h_1(c)\over \rho_s^2\, 	r_s^3} 
	= \langle\sigma  v\rangle_f 
	{f_2(a)\,h_1(c)\over f_1(a)\,h_2(c)}
\label{slow_dec_sv}
\ee 
from (\ref{gamma_eff}) and (\ref{fneq}), that the cross
sections increase to $970\times 10^{-22}$ and $370\times
10^{-22}\,$cm$^3\,$s$^{-1}$ respectively, for the two halo 
profiles.  In table \ref{tab1}
we give the ranges including the uncertainty in the flux
$F_\gamma$.

We have also explored the effect of a cored versus cuspy DM halo.
Ref.\ \cite{Tempel:2007eu} fits the M31 DM halo to several
profiles including NFW (cuspy) and Burkert (cored), the latter
being given by
\be
	\rho_{\rm B}(r) = {\rho_s\over (1 + r/r_s)(1 + r^2/r_s^2)}
\label{burk}
\ee
with best fit values  $r_s = 6.86\,$kpc, $\rho_s = 5.72\times 10^7\,
M_\odot/$kpc$^3$ for Burkert.  The LOS integral of $\rho^2$ over the
0.22$^\circ$ FOV differs only by a factor of 1.1-1.7 relative to the
NFW profiles considered above, and so
we conclude that the cusp versus core issue is not a great source of
uncertainty for M31.  We will see however that it is much more
important for the Milky Way.

\section{The Milky Way}  Ref.\ \cite{Riemer-Sorensen:2014yda} obtains
upper limits on the flux of X-ray lines in the energy intervals
3-6 and 2-9 keV, based upon Chandra observations of a $16'\times 16'$
region at the galactic center, with the central disk of angular
size $2.5'$ excised.  For simplicity we model this region by an
annulus of equivalent area with $2.5'<\theta<9'$.  The limits on 
the flux at photon energy 3.55 keV are found to be 12.1 and 
$21.4\times 10^{-6}$ counts/cm$^2$/s respectively for the two energy
intervals (see footnote \ref{rsr}).

\begin{figure*}[t]
\centerline{\includegraphics[width=\columnwidth]{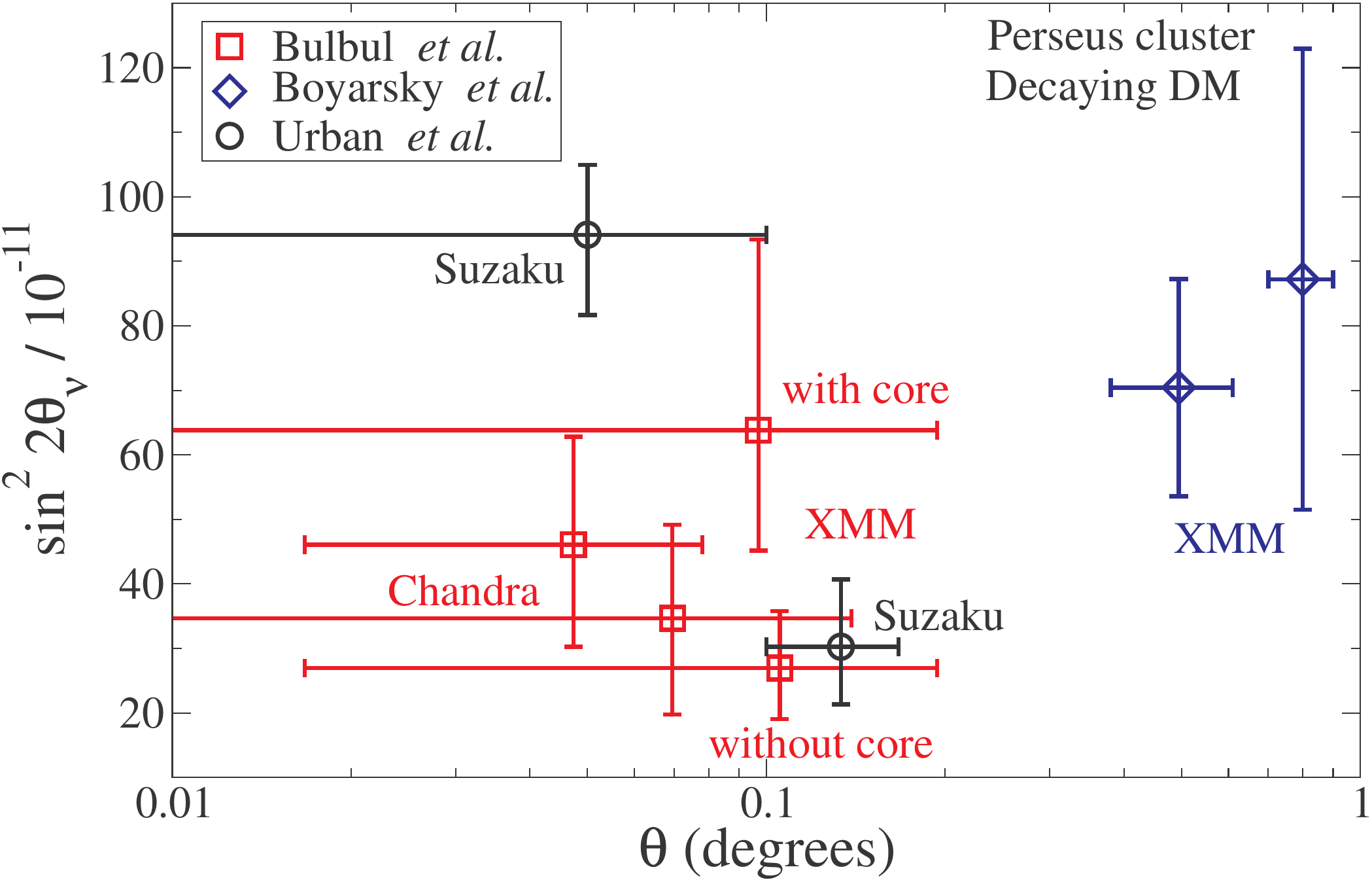}
\includegraphics[width=\columnwidth]{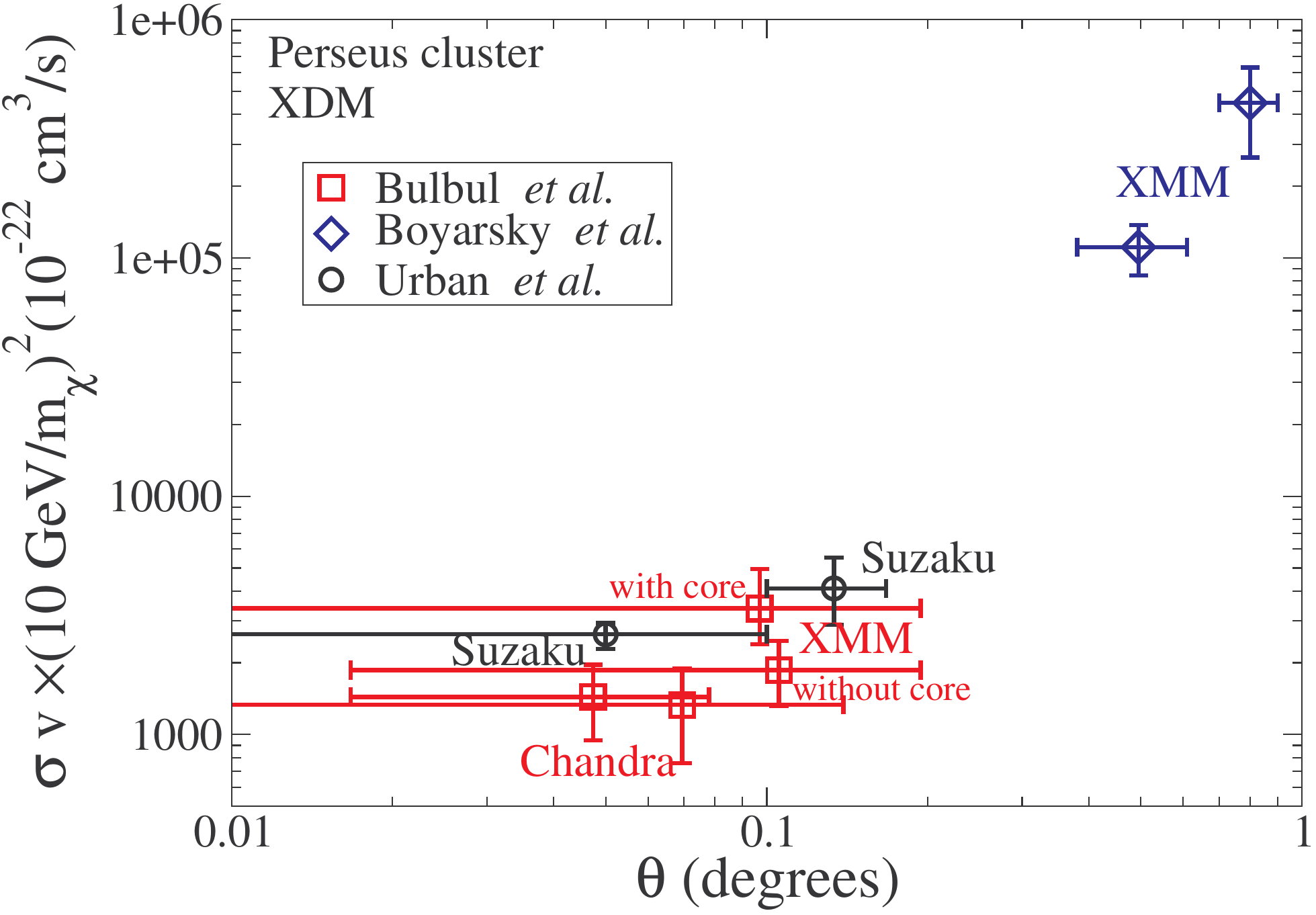}}
\caption{Left: mixing angle of decaying sterile neutrino that would
give the measured fluxes for Perseus in the angular intervals
indicated by horizontal error bars.  Vertical bars correspond to
errors in fluxes.  Right: similar to left, but showing the
cross section for excited dark matter models.  Perseus data alone
would favor the decaying DM scenario, or possibly XDM with slow decays of the 
excited state.}
\label{pers_fig1}
\end{figure*} 

We apply these flux limits directly to find the corresponding  limits
on $\Gamma_\nu$ and 
$\langle\sigma v\rangle$ using eqs.\ (\ref{fluxeq1},\ref{fluxeq2}). 
To explore the
dependence upon the assumed Milky Way DM profile, we first
compute $\int
d^{\,3}x\,\rho^n/x^2$ for a range of NFW profiles suggested by
simulations (see ref.\ \cite{read}), with $r_s = 19{+7.5\atop -5.4}\,$
kpc at 68\% confidence, and varying the local density $\rho_\odot$ between
$(0.3-0.4)$\,GeV/cm$^3$ while keeping the distance $r_\odot$ to the GC 
fixed at $8.3\,$kpc. (The effect of a cored profile will be considered
below.)  Varying $r_s$, $\rho_s$ and the limit on the flux
leads to the range of upper limits the neutrino mixing angle
$\sin^2 2\theta < (6-20)\times 10^{-11}$ and on the cross section
$\langle\sigma v\rangle (10{\rm\, GeV}/m_\chi)^2 < (0.15-1.1)$
in units of $10^{-22}$cm$^3$/s, for the case of fast decays.  Repeating
this for the $1\sigma$ range of Milky Way NFW profiles found in 
ref.\ \cite{Nesti:2013uwa}, we get the smaller range of $\langle\sigma
v\rangle(10{\rm\, GeV}/m_\chi)^2 < (0.07-0.15)$.  The larger range
is shown in table \ref{tab1}.

On the other hand, ref.\ \cite{Boyarsky:2014ska} finds positive
evidence for the line from the inner $14'$ of the GC using XMM-Newton data, with
a flux of $(29\pm 5)\times 10^{-6}$ counts/cm$^2$/s which is
consistent with the previous bounds at the $2\sigma$ level.  Following
the same procedure as for the upper limits, we find the allowed
ranges of the cross section to be $(0.11-0.65)\times 10^{-22}$cm$^3$/s
for the DM profiles reported in ref.\ \cite{read}.

For slowly decaying excited states, again using (\ref{slow_dec_sv})
we find that the range of upper limits from
\cite{Riemer-Sorensen:2014yda} becomes 
$(0.19-1.7)\times 10^{-19}$cm$^3$/s, while the
range of measured values from \cite{Boyarsky:2014ska} is 
$(0.39-2.7)\times 10^{-19}$cm$^3$/s.  These are much larger than
the corresponding M31 values because of the large factor
$f_2|^{a_2}_{a_1}/f_1|^{a_2}_{a_1}=320$ for the MW in eq.\ 
(\ref{slow_dec_sv}), compared to $\sim 1$ for M31.

Because the FOV is so strongly concentrated on the galactic center for
the MW observations,
the model predictions are extremely sensitive to the assumed behavior
of the DM profile in this region.  Ref.\ \cite{Nesti:2013uwa} prefers
the Burkert profile (\ref{burk}) as the best fit to the MW.  The LOS
integral of $\rho^2$ is much smaller with the best fit cored profiles
shown in fig.\ 3 of \cite{Nesti:2013uwa} than the corresponding NFW
profiles also fit there, by a factor of 500-1000.  In table \ref{tab1}
we give the ranges of cross sections for fast decays in these cored
profiles as well as in the NFW profiles.  These are extreme cases, and
one could expect the true profile to give cross section values
somewhere in between.  Because of this uncertainty, the MW
observations might be considered not very constraining when trying to
distinguish between different kinds of DM models.

\section{Perseus Cluster}  Ref.\
\cite{Bulbul:2014sua} obtains several flux measurements of the line,
centered on the Perseus cluster using XMM and Chandra data. In two of
them, the   central region of radius $1'$ is removed because of large
X-ray  fluxes possibly having an origin from atomic transitions in the
core of the cluster.  These have fluxes of  $21.4{+7.0\atop-6.3}$ with
XMM MOS and  $10.2{+3.7\atop-3.5}$ with Chandra ACIS-S, in units of
$10^{-6}$cm$^{-2}$s$^{-1}$, and  respective fields of view $700''$ and 
$281''$ (approximating the square by a circle of equivalent area) in
radius, minus the excised central region.  There are also two
measurements including the core, with fluxes of $52.0{+24.1\atop-15.2}$
from XMM MOS and 
$18.6{+7.8\atop-8.0}$ from Chandra ACIS-I.  More recently ref.\
\cite{Urban:2014yda} measured nonzero fluxes using Suzaku data,
with $F_\gamma = 2.87{+0.33\atop -0.38}\times 10^{-7}$cm$^{-2}$s$^{-1}$arcmin$^{-2}
\times 36\pi\,$arcmin$^{2}$ from the central $0.1^\circ$, and 
$F_\gamma = 4.78{+1.65\atop-1.41}\times
10^{-8}$cm$^{-2}$s$^{-1}\times (18^2-36\pi)$ from a surrounding
region going out to $0.17^\circ$.
We model the fields of view as annular regions of equivalent area.

To evaluate the LOS integrals, we use the DM density profile
determined by  ref.\ \cite{Simionescu:2011ii}.  It  finds virial mass
and radius $M_{200} = 6.65{+0.43\atop -0.46}\times 10^{14}M_\odot$ and
$R_{200} = 1.79\pm 0.04\,$Mpc, and concentration parameter 
$c_{200}=5.0\pm
0.5$,\footnote{The value $c_{200}=5$ corresponds to $c_{500}=3.3$, so this is
consistent with our previous assumptions about NFW parameters of
clusters for the stacked analysis} leading to NFW scale radius $r_s = R_{200}/c = 0.36\pm
0.04\,$Mpc, and density $\rho_s = M_{200}\, c^3/(4\pi\, R_{200}^3\,
h_1(c)) =  (1.2\pm 0.3)\times 10^{15}M_\odot/$Mpc$^3$, where the values
are anticorrelated with those of $r_s$ via the dependence upon $c$ and
we ignore the smaller error due to $M_{200}$.  We take the distance to
the cluster to be $d=74\,$Mpc.

The data divided by the LOS integrals in (\ref{fluxeq1}) or
(\ref{fluxeq2}) are plotted
in fig.\ \ref{pers_fig1}, so that consistency with a given model
should result in no dependence on the angle from the center.
 The data at low angles do not show a clear
preference for decaying versus annihilating or scattering DM models,
but including those at larger off-center angles would disfavor 
scattering (XDM) relative to decays.
Ref.\ \cite{Boyarsky:2014jta} reports 
X-ray fluxes of $(13.8\pm 3.3)$ and  $(8.3\pm 3.4)\times
10^{-6}$/cm$^2$/s in two off-center 
angular bins with $\theta =(0.38^\circ-0.61^\circ)$ and 
$(0.7^\circ-0.9^\circ)$ respectively, and with average field of view
537 arcmin$^2$.
We model the fields of view by part of an annular region bounded by
the polar angles $\theta_i$ given above, and an azimuthal interval
$\delta\phi$ such that $\delta\phi(\cos\theta_2-\cos\theta_1) = 537$ 
arcmin$^2$.  Then $\delta\phi = 75.1^\circ,\, 53.4^\circ$ respectively
for the two bins, and we can take $\delta\phi/360^\circ$ times 
the formula (\ref{fneq}) for the density integrals when applying
(\ref{fluxeq2}) or (\ref{slow_dec_sv}). (For annular regions bounded
by two polar angles, one must replace $f_2(a)/f_1(a) \to
f_2|^{a_2}_{a_1}/f_1|^{a_2}_{a_1}$.) 

Although the Perseus data by themselves appear to be more consistent
with decaying DM than with XDM, one should keep in mind that the
mixing angle required by the combined Perseus observations is
in tension with limits from the Coma cluster, dwarf spheroidals and stacked galaxies
to be discussed below, suggesting that the DM signal, if present, may
be contaminated by other backgrounds.\footnote{For an alternative
explanation, see ref.\ \cite{Cicoli:2014bfa}}\    Ref.\ 
\cite{Urban:2014yda}
notes that the evidence for the line in Perseus disappears if a 
sufficiently complex model of atomic line backgrounds is adopted.
Moreover, the large signal at large off-center angles might be
consistent with the hypothesis that baryons in clusters tend to 
be concentrated at the outskirts due to transport by shock waves
\cite{Rasheed:2010pq}.
On the other hand,
the XDM model can reconcile some of the Perseus data with the upper
limits.   In our XDM fits to the data, we will treat the large-angle
observations and those including the core as outliers and retain only
the lower-angle noncore fluxes reported by \cite{Bulbul:2014sua}. 
Including the data of ref.\ \cite{Urban:2014yda} deteriorates the
quality of the fits somewhat but does not change the shape of the allowed
regions significantly.

\section{Other nearby clusters}
Ref.\ \cite{Bulbul:2014sua} finds a positive signal from combining
spectra of nearby clusters Coma, Centaurus and Ophiuchus (CCO), with
averaged flux $F_\gamma = 15.9{+3.4\atop-3.8}\times 10^{-6}/$cm$^2/$s over the XMM MOS $700''$ field of view.
The respective weighting factors for averaging the
spectra are $0.35,\,0.51,\,0.14$ \cite{bulbul}.  In addition, 
a 90\% confidence level upper limit of $F_\gamma < 9.1\times 10^{-6}/$cm$^2/$s on the flux from the 
Virgo cluster was found using Chandra ACIS-I with $500''$ FOV.
More recently,
ref.\ \cite{Urban:2014yda} obtained 95\% c.l.\ 
upper limits on the line fluxes
from Suzaku spectra of Coma, Virgo and Ophiuchus, with that from 
Coma giving the most stringent constraints on dark matter models: 
$F_\gamma < 2.65\times
10^{-9}/$cm$^2/s$/arcmin$^2\times (18 {\rm\ arcmin})^2$.

To translate these into DM model constraints we follow the procedure
of ref.\ \cite{Urban:2014yda} for determining the NFW halo parameters,
similar to the one which we employed for the stacked clusters:
a universal concentration $c_{200}=4.1$ is
assumed, implying $\rho_s = \delta_c\rho_{\rm crit} = 7\times
10^{14}M_\odot/$Mpc$^3$, while the scale radius is given in terms of
virial masses as $r_s^3 = M_{200}/(4\pi\rho_s h_1(c))$.  As in 
\cite{Urban:2014yda} we take
$M_{200} = (8.54,\, 1.40,\,1.47)\times 10^{14}M_\odot$ respectively 
for Coma, Virgo and Ophiuchus, while for Centaurus we take
$M_{200} = 3\times 10^{14}M_\odot$, which follows from the average
temperature $kT = 3.68$ keV \cite{Horner:1999wf} and the $M_{200}$-$T$ scaling
relation of \cite{Arnaud:2005ur}.  

Using these parameters and eqs.\ (\ref{fluxeq1}-\ref{fluxeq2}),
we find from the CCO observation the values for the mixing angle and
cross section given in table \ref{tab1} (with uncertainties due to the
halo profile estimated by varying $c=3-4$), roughly compatible with the
values obtained by the same authors for the Perseus cluster.  The
upper limits from Virgo are also marginally compatible with these
values.  However the limit on the mixing angle from the Coma cluster
flux limit of ref.\ \cite{Urban:2014yda} is quite stringent, 
$\sin^2 2\theta_\nu \lesssim 2\times 10^{-11}$, and at odds with the
values indicated by the other positive observations of the line.
The limit on the XDM cross section from Coma on the other hand is
in a lesser degree of tension, $\langle\sigma v\rangle_f < 200\times
10^{-22}$cm$^3$/s for $m_\chi=10$ GeV, only marginally below the
range of values indicated by the stacked clusters.

\section{Dwarf spheroidal limits}  Malyshev {\it et al.}\  
\cite{Malyshev:2014xqa} give limits
on the neutrino mixing angle from XMM-Newton observations of eight
dwarf spheroidal galaxies in the Milky Way.  In their analysis,
no assumption was made as to the shape of the dwarf density profiles,
since they approximate $\int d^{\,3}x\,\rho/x^2 = M_{FOV}/d^2$, where
$M_{FOV}$ is the DM mass within the FOV and $d$ is the distance to the
dwarf spheroidal. This enclosed mass was estimated
in a way that was assumed to be relatively independent of the
particular profile.  

However for DM scattering, the signal goes like $\rho^2$ and can thus
have a stronger dependence on the profile shape.  Here we consider the
possibilities that the dwarf density profiles are either NFW as in
eq.\ (\ref{NFW}) or cored, with $\rho_{\rm cored} =
\rho_s/(1+r/r_s)^3$. Following ref.\ \cite{Walker:2009zp} that 
presents evidence for universal density profiles for dwarf spheroidal
galaxies, we take $r_s = 0.795\,$kpc for each NFW profile, and $r_s =
0.150\,$kpc for cored, deriving the corresponding XDM limits on
$\langle\sigma v\rangle$ for both cases.  

Moreover
ref.\ \cite{Malyshev:2014xqa} accounts for the flux from the Milky Way 
halo, for which the profile is taken to be NFW, but with two possible
values of the scale radius and normalization, \cite{klyp}.  These
are $r_s = 21.5\,$kpc, $\rho_s = 4.9\times 10^6 M_\odot/$kpc$^3$ and
$r_s = 46\,$kpc, $\rho_s = 0.6\times 10^6 M_\odot/$kpc$^3$,
respectively, and referred to as the ``mean'' and ``minimal'' 
halo models.\footnote{Ref.\ \cite{klyp} does not consider the
minimal model to provide a good fit to properties of the MW, but 
here it
illustrates the impact of an unrealistically low DM halo density 
on the dwarf constraints.} 
The total
DM density is thus $\rho = \rho_s + \rho_{MW}$ in each observed
dwarf field of view (FOV), complicating the evaluation of the 
integral of $\rho^2/x^2$,  which we carry out numerically. 
(Analytic expressions for the contributions from the dwarf halo
densities $\rho_s$ alone are given in the appendix, and were used
to check the numerics.) 
The angular radius of the FOV is taken to be the minimum of
$\theta_{1/2} = r_{1/2}/d_i$ and the XMM-Newton FOV, $15'$.

Ref.\ \cite{Malyshev:2014xqa} tabulates the distance to each dwarf,
its half-light radius $r_{1/2}$ (where the intensity profile of its visible
light output drops by a factor of 2 relative to the center), and the
mass $M_{1/2}$ enclosed within $r_{1/2}$.  These are sufficient for
determining the parameters $\rho_s$ for either the NFW or
the cored profile each dwarf, given the assumed values of 
$r_s$ mentioned above.  The only further information
required for computing $\int d^{\,3}x\, \rho^2/x^2$ is the angle
$\phi_i$ between the line of sight (LOS) to the dwarf and that to the galactic
center, since the DM density of the Milky Way halo
along the LOS to the dwarf depends upon $\phi_i$.  This angle is
related to the galactic coordinates $(b,l)$ of the dwarf by $\cos\phi = 
\cos(b)\cos(l)$.  We find that $\cos\phi_i = \{-0.159,\,0.052,\,
-0.224,\,-0.455,\,-0.183,\,-0.704,\,-0.510,$ $-0.496\}$ respectively
for the satellites Carina, Draco, Fornax, Leo I, Ursa Minor, 
Ursa Major II, Willman I, and NGC 185.  The distance to the GC
is taken to be 8.5 kpc for consistency with  \cite{Malyshev:2014xqa}.

The expected fluxes for the decaying neutrino model are given in 
\cite{Malyshev:2014xqa}, so we need not recompute $\int
d^{\,3}x\,\rho/x^2$ in (\ref{fluxeq1}).   It can be deduced from the
fluxes using the limit on the decay rate $\Gamma_\nu < 6.68\times
10^{-29}\,$s$^{-1}$ that we infer from the limit on the mixing angle $\sin^2
2\theta < 2.67\times 10^{-11}$ for the mean MW model. These limits are
weaker by a factor of 1.78 for the minimal MW model. Carrying out the
$\rho^2$ integrals and weighting them by the exposures given in 
\cite{Malyshev:2014xqa}, we find the following limits on 
$\langle\sigma v\rangle (10{\rm\, GeV}/m_\chi)^2$, in units of
$10^{-22}$cm$^3$/s: for the NFW dwarf profiles, 0.18 and 0.26
respectively, in the mean and minimal MW halo models; for the cored
dwarf profiles, we obtain the same values to two significant figures,
thus finding little difference between cored versus NFW profiles,
although there is some dependence upon the assumed shape of
the MW halo.

\section{Limit from stacked galaxy spectra}
\label{galaxies} Ref.\ 
\cite{Anderson:2014tza} finds a limit of $\sin^2 2\theta < 0.47\times
10^{-10}$ for a decaying neutrino with $m_\nu=7.1$ keV,
from stacking Chandra spectra of 81 galaxies.  The authors obtain a
somewhat stronger limit of $\sin^2 2\theta < 0.19\times
10^{-10}$ from XMM-Newton data for 89 galaxies.  They assume NFW profiles for which the scale radii are
related to the virial radius as $r_s = R_{\rm vir}/c_{200}$, where the
concentration parameter is related to the virial mass $M_{\rm vir}$
as determined by ref.\ \cite{Prada:2011jf}.  The relation can be
fit by $\log_{10} c_{200} = 1.85 - 0.08 \log_{10} M_{\rm vir}$ (where
$M_{\rm vir} = M_{200}$ is in units of $h^{-1} M_\odot$) for 
$\log_{10}M_{\rm vir}<14.6$, and remaining constant for higher masses.
Ref.\ \cite{Anderson:2014tza} tabulates $R_{\rm vir}$ and 
$M_{\rm vir}$ (also called $M_{\rm halo}$ in that reference) as well
as the distances to each galaxy.  This allows us to 
construct the 
NFW profiles for their galaxies (using $\rho_s = M_{\rm vir}/(4\pi
r_s^3 h_1(c))$) to calculate the quantities in
eq.\ (\ref{fluxeq1}-\ref{fluxeq2}).

To avoid sensitivity to the central cusp of the NFW distribution
(which may not be present in realistic simulations accounting for
effects of baryons),
the signal is taken between $r=0.01$ and $1$ times $R_{\rm vir}$
of each galaxy.  Analytic expressions can be found for the LOS
integrals for NFW density (and density-squared) profiles 
(see appendix).   We compute the exposure-time-weighted averages
of the density integrals separately for the Chandra and XMM-Newton
data to obtain the equivalent limits on 
$\langle\sigma v\rangle (10{\rm\, GeV}/m_\chi)^2$ for the two data
sets.  The limit on $\langle\sigma
v\rangle_f$ is related to that on the decay rate by $\Gamma_\nu (m_\chi^2/m_\nu) \langle M_{\rm
vir}f_1|^c_{0.01c} d^{-2} h_1^{-1}(c)\rangle/$ $\langle M_{\rm
vir}\rho_s f_2|^c_{0.01c}d^{-2} h_1^{-1}(c)\rangle$ for fast decays,
and for slow decays the relation is $\langle\sigma
v\rangle_s = 
$ $\Gamma_\nu (m_\chi^2/m_\nu)\langle(h_2/h_1)\rho_s\rangle^{-1}$.
From table \ref{tab1} we see that the resulting limits on XDM are less
constraining than the claimed observations from M31 or MW, and so we
can omit this constraint from the fits we undertake next.

\section{Compatibility of XDM with data}  
\label{compat} Decaying dark matter is
ostensibly at odds with the required values of the lifetime (here
parametrized by the sterile neutrino mixing angle) for the claimed
observations, versus the upper limits from null searches.  
Superficially, it would appear that the same is true for annihilating
DM models or XDM, from the required values versus upper limits for the 
annihilation or excitation cross section.  However for XDM there is an
additional parameter, since the 
cross section depends upon the DM velocity due to the energy
threshold needed to create the excited states.  In the center-of-mass
frame, this corresponds to the relative velocity threshold
\be
	{v_t/c} = \sqrt{8\, \delta m_\chi/m_\chi}
\label{vteq}
\ee
The kinetic energy of the scattering DM particle is
 $m_\chi (v_t/2)^2/2= \delta m_\chi$, necessary for creating the
excited state with mass $m_\chi + \delta m_\chi$.   This can be used to
explain why the X-ray signal from XDM would be stronger in systems
with larger DM velocities.   
 
Indeed, the phase space
averaged cross section depends upon the velocity dispersion of 
the DM, $\sigma_v = \langle v^2\rangle^{1/2} = \sqrt{3/2}\,v_0$ (where
$v_0$ is the circular velocity), 
corresponding to an assumed Maxwellian distribution $f(v)=
Ne^{-(v/v_0)^2}$.  The averaged cross section can be approximated 
as 
\bea
\label{xsecteq}
	 \langle\sigma v\rangle &=& \sigma_0 v_t \gamma,\\
\label{gamma}
	\gamma &=& \left\langle\sqrt{v_{\rm rel}^2/v_t^2-1}\,\,\Theta(v_{\rm rel}-v_t)
	\right\rangle,
\eea
and $\langle F(v_{\rm rel}) \rangle = \int d^3 v_1 d^3 v_2 f(v_1)
f(v_2) F(|\vec v_1-\vec v_2|)$,   
assuming that there is no significant velocity-dependence in the cross
section other than that coming from the phase space integral.
The functional form of $\gamma(v_0/v_t)$ is plotted in 
ref.\ \cite{Cline:2014kaa} (and will appear in our figures showing
fits to the data below).  For $v_0>v_t$, it is approximately
linear, while for $v_0<v_t$ it falls
exponentially, 
\be	
\label{gamma_eq}
	 \gamma \cong 1.4\left\{\begin{array}{ll} 
	(v_0/v_t),& v_0\gtrsim 1.3\,v_t\\
	e^{-(v_t/v_0)^2/2},& v_0\lesssim 0.3\,v_t \end{array}
	\right.
\ee
For $v_0\sim
v_t$ there is intermediate behavior that smoothly connects these 
two expressions.\footnote{The analytic ansatz
$\gamma\cong 1.4 (v_0/v_t)f_+ +  1.4\,e^{-(v_t/v_0)^2/2})f_-$
provides a good fit, where $f_\pm =
[1\pm\tanh(20\log_{10}(v_0/v_t)]/2$.}

This additional dependence can explain why no X-ray signal is seen
from dwarf ellipsoidal galaxies, whereas it is strong enough in galaxy
clusters and nondwarf galaxies.  However, it cannot explain the 
fact that the Perseus cluster seems to give a much stronger line 
relative to other galaxy
clusters, especially at large off-center angles.  To pursue the DM
explanation,  we must assume that these large-angle Perseus
observations are contaminated by some  background, as was suggested by
ref.\ \cite{Bulbul:2014sua}, or else that only the Perseus
observations are indicative of DM decays, and that the other positive
claims are somehow spurious.  On the other hand, the error bars on the
on-center Perseus measurements are sufficiently large that they do not
greatly diminish the goodness of our fits if we include them. In the
following, we will show results of fits in which  Perseus fluxes are
treated as outliers that require  further explanation, but we will
also indicate the effect of the on-center Perseus fluxes  in the
computation of the $\chi^2$, using the Bulbul {\it et al.}\ values
shown in table \ref{tab1} for the required value of the cross section.

To estimate the average velocity dispersion $\sigma_v$ for clusters,
we have
identified $\sigma_v$ for 22 out of the 73 clusters
studied by Bulbul {\it et al.}\ using ref.\ \cite{struble}; see
table \ref{tab2}.  Their average $\sigma_v$ is 1055 km/s, and their
average weighted by the exposures of ref.\ \cite{Bulbul:2014sua} 
is 975 km/s.

\begin{table}[htb]
\begin{tabular}{|l|l||l|l||l|l||l|l|}
\hline
name & $\sigma_v$ & name & $\sigma_v$ & name & $\sigma_v$
& name & $\sigma_v$ \\
\hline
Perseus & 1282 & A262 & 588 & A478 & 904 & A496 & 714 \\
\hline
A665 & 1201 & A754 & 931 & A963 & 1350 & A1060 & 647\\
\hline
 A1689 & 1989 &A2063 & 659 & A2147 & 821 & A2218 & 1370 \\
\hline
 A2319 & 1770 & A2811 & 695 & A3112 & 950 & A3558 & 977\\
\hline
A2390 & 1686  &  A3571 & 988 & A3581 & 577 & A3888 & 1831\\
\hline
A4038 & 882 & A4059 & 628 & & & & \\
\hline
\end{tabular}
\caption{Clusters from ref.\ \cite{Bulbul:2014sua} whose velocity dispersions
$\sigma_v$ (in km/s) are tabulated in ref.\ \cite{struble}.}
\label{tab2}
\end{table}

\begin{figure}[t]
\centerline{\includegraphics[width=\columnwidth]{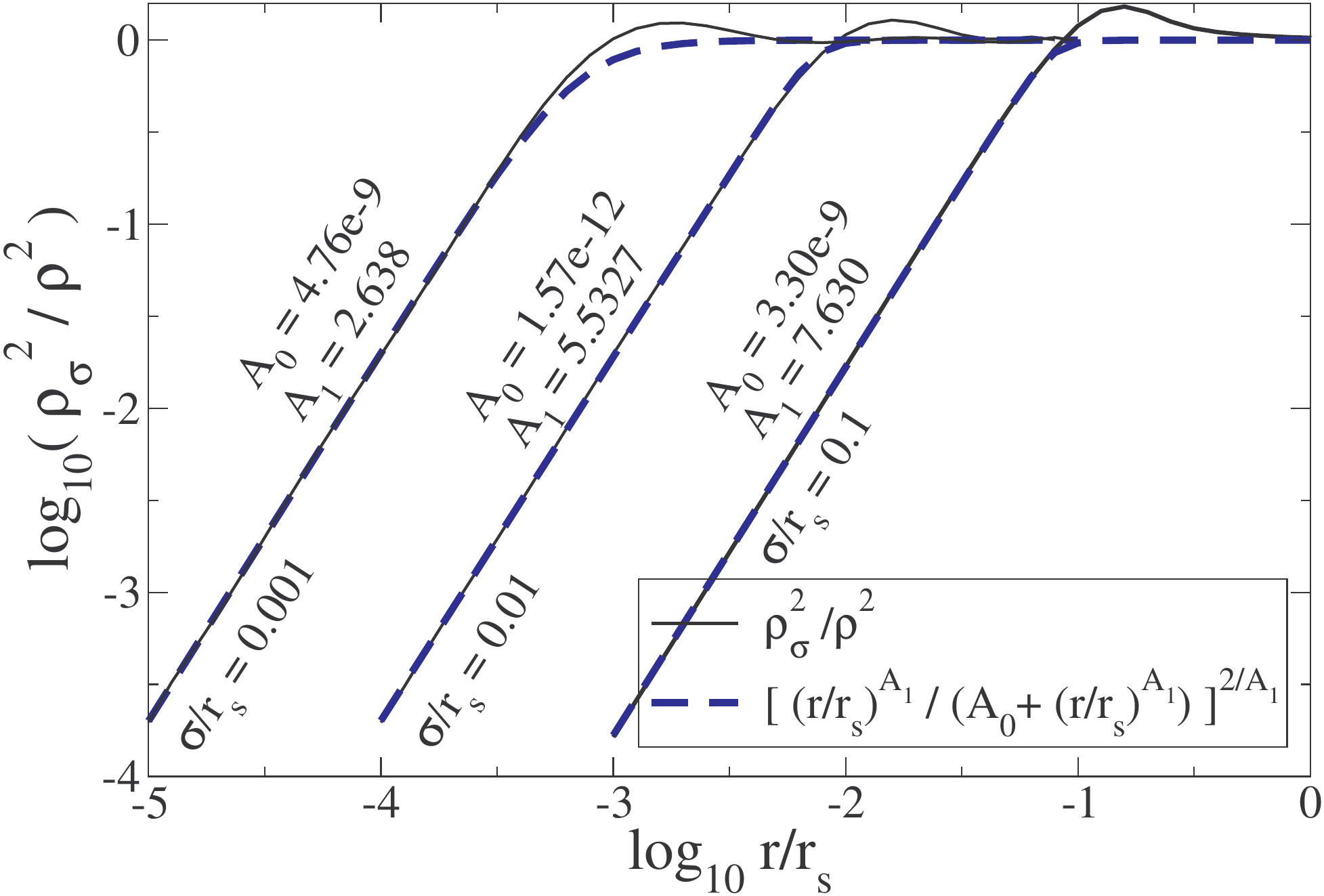}}
\caption{Solid curves: ratio of smeared $\rho^2$ to original $\rho^2$, eq.\
(\ref{smeared}), for NFW profile with smearing lengths $\sigma = 
(0.1,\,0.01,\,0.001)\,r_s$ (from right to left), 
and approximate analytic fits (dashed curves).}
\label{corr-fact}
\end{figure}

\subsection{M31 versus Milky Way}
\label{M31sect}

We have seen that for NFW DM profiles, there is a discrepancy
between the X-ray line strengths from M31 and from the Milky Way, which require  seemingly
incompatible values of the cross section.  Basically the claimed
signal from the very small
field of view of the MW (relative to its virial radius) should be
much higher for a NFW profile, to be compatible with that from M31.
This suggests that the density of decaying excited states is lower in 
this central region than predicted.  We have identified two
ways of addressing this: (1) the lifetime of the excited state is
long enough for the particles to stream out of the central region
before decaying; (2) the halo profile is less cuspy than NFW.
We will consider both of these possibilities in the following.

\subsubsection{Intermediate lifetime of excited state}

Table \ref{tab1} indicates a dramatic increase in the required scattering
cross sections for the case of slow decays versus fast ones
suggesting that there exist 
some intermediate values of the lifetime where the XDM interpretation
of M31 and
MW fluxes could be made compatible.  It is important to realize that
the MW observation has a FOV that covers only the inner part of the 
halo with $r/r_s <  10^{-3}$, whereas the M31 observation covers a much
larger fraction $r/r_s < 0.13$.  This implies that a relatively short
excited state lifetime $\tau$ (compared to the dynamical timescale for
galaxies) could be enough to deplete the inner
region of  the MW due to DM transport during time $\tau$, reducing the
signal there and boosting the required value of $\langle\sigma
v\rangle$, whereas it would have a small effect on the observed region
of M31.

\begin{figure}[t]
\centerline{\includegraphics[width=\columnwidth]{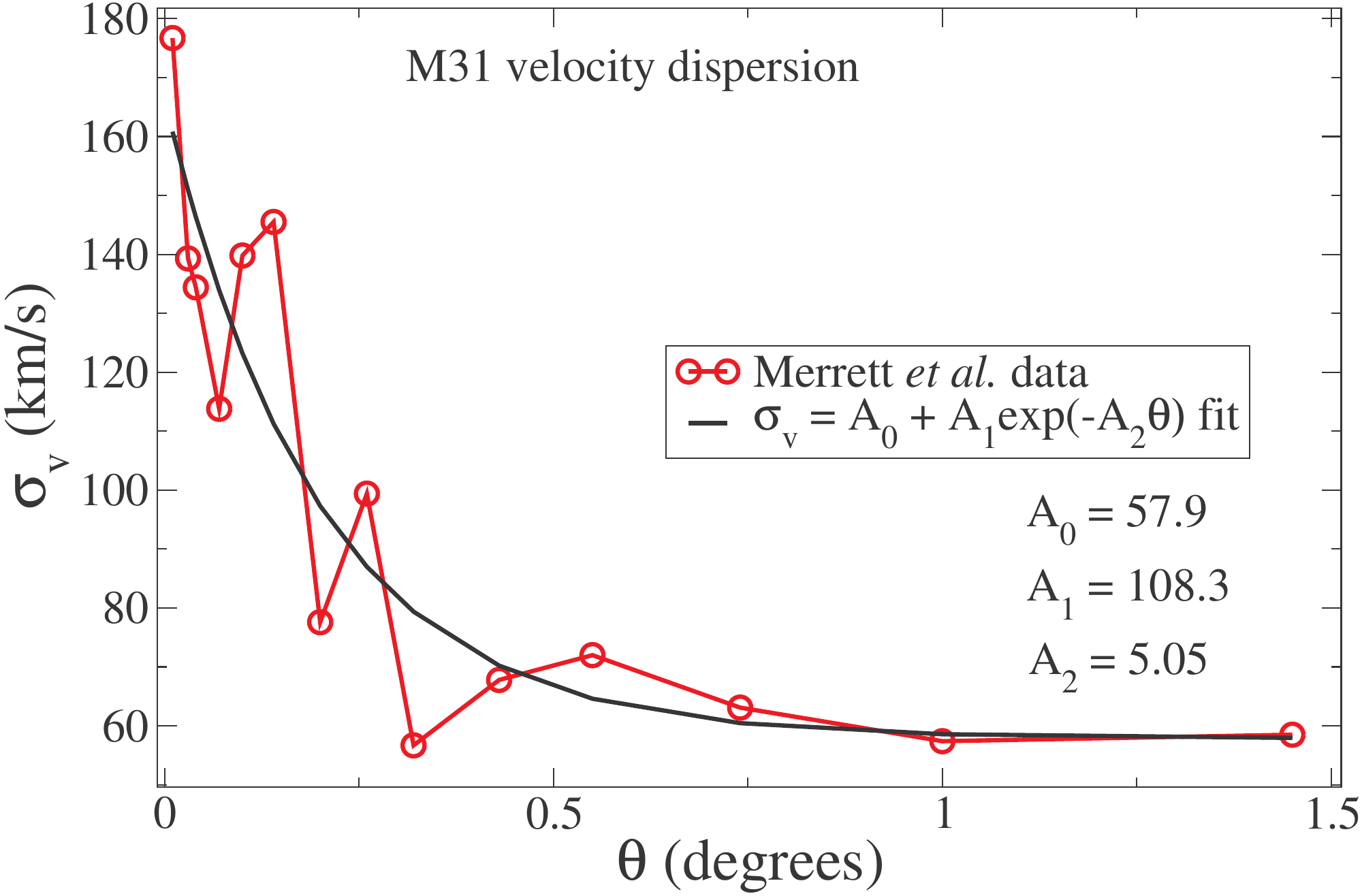}}
\caption{Measurements of M31 velocity dispersion from ref.\ 
\cite{merrett}, and an analytic fit to the data.}
\label{M31-fig}
\end{figure}

\begin{figure}[t]
\centerline{\includegraphics[width=\columnwidth]{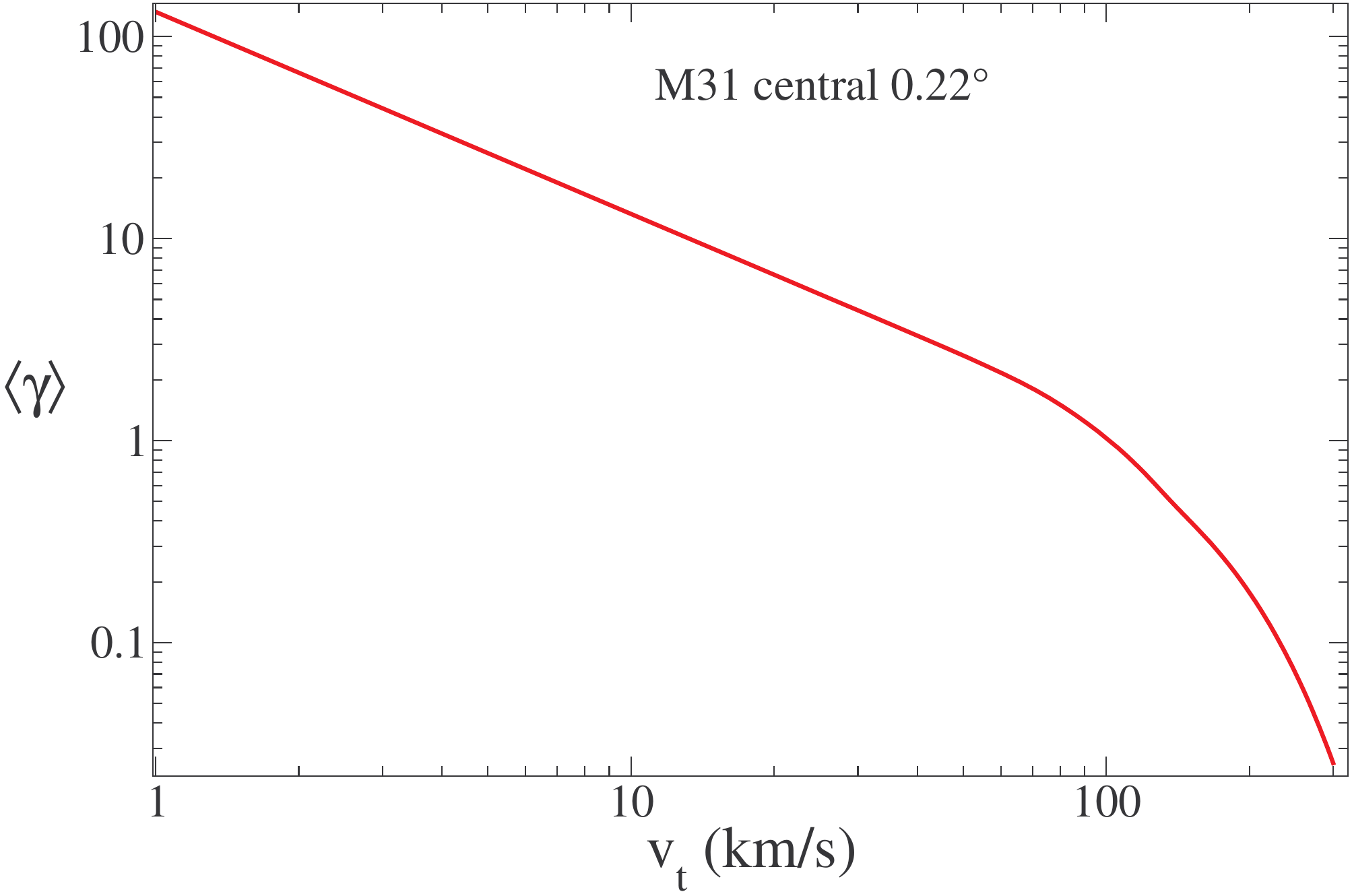}}
\caption{Average value of 
$\gamma$, eq.\ (\ref{gamma}), $\langle\gamma\rangle = 
[\int d^{\,3}x\, \gamma\rho^2/x^2]/[\int d^{\,3}x\, \rho^2/x^2]$,  
integrated over the $1.5^\circ$ FOV of the M31 observations in
ref.\ \cite{Boyarsky:2014jta}.}
\label{M31-gamma}
\end{figure}

To model the effect of a relatively short excited state lifetime,
we will assume that the flux formula (\ref{fluxeq2}) holds, but with
$\rho^2$ replaced by a smeared version,
\be	
	\rho^2(x) \to \rho^2_\sigma = N\int d^{\,3}x'\,
	e^{-(\vec x - \vec x')^2/\sigma^2}\, \rho^2(x')
\label{smeared}
\ee
where $\sigma$ is the streaming length of the excited states before
they decay, $\sigma \sim \sigma_v \tau$, with $\sigma_v$ being the
velocity dispersion.  We consider the relevant values $\sigma = 
(10^{-3},\,10^{-2},\,10^{-1})\, r_s$ 
(taking for simplicity $r_s = 23.8\,$kpc for both M31 and
MW).  By numerically evaluating the integral in (\ref{smeared}) for
an NFW density profile, we find that $\rho^2_\sigma$ is given by
a function that can be approximated as
\be
	{\rho_\sigma^2 \over \rho^2}\cong \left( (r/r_s)^{A_1}\over
	A_0 + (r/r_s)^{A_1}\right)^{2/A_1}
\ee
where the values of $A_{0,1}$ for a given $\sigma/r_s$ are
shown in fig. \ref{corr-fact}.  We see that the
$1/r^2$ behavior in $\rho^2$ gets canceled out in the smeared profile
below distances $r\lesssim \sigma$.  Hence the flux is reduced for
a field of view subtending this region, while it remains relatively
unchanged for a much larger FOV.  The exact result
is compared to this approximate analytic fit in 
figure \ref{corr-fact}.

\begin{figure*}[t]
\centerline{\includegraphics[width=2.2\columnwidth]{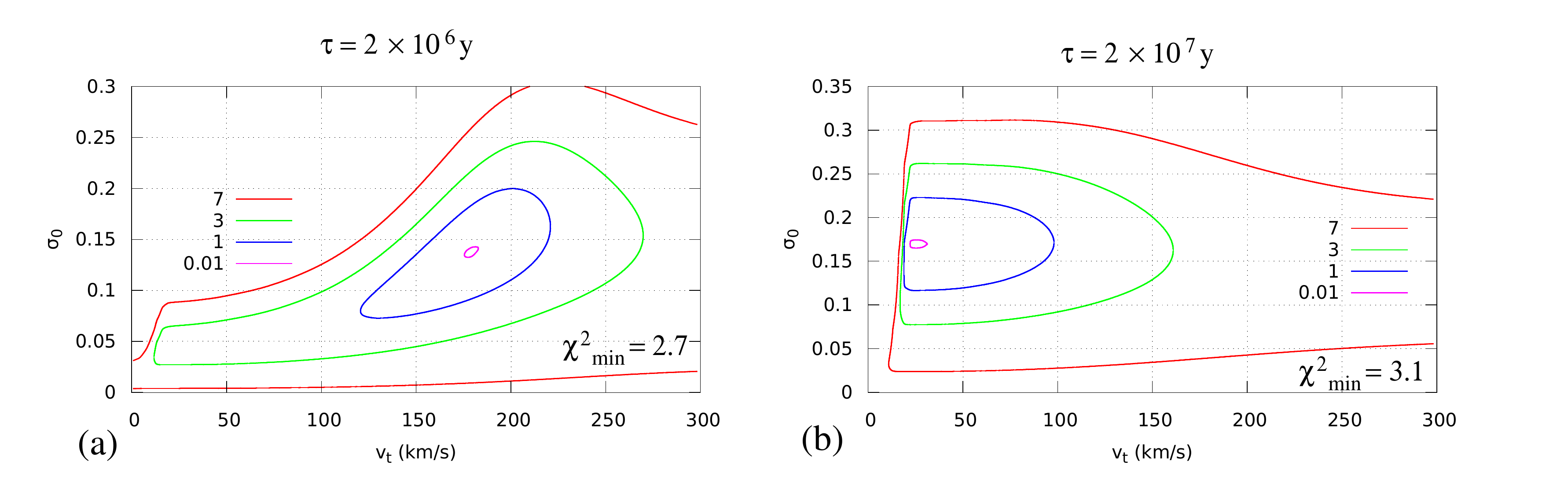}}
\caption{Contours of $\delta\chi^2$ in the $v_t$-$\sigma_0$ plane for excited dark matter with lifetime
$\tau\sim 2\times 10^6\,$y (a) and $2\times 10^7\,$y (b), excluding
Perseus data.  Minimum values of $\chi^2$ increase by $\sim 3$ but
shapes of contours do not change significantly with inclusion 
of Perseus cluster.
$\sigma_0$ is in units of $10^{-22}$cm$^3$s$^{-1}/$(km/s)$\cdot (m_\chi/{\rm 10\
GeV})^2$. }
\label{cont-panel}
\end{figure*}

\begin{figure*}[t]
\centerline{\includegraphics[width=2\columnwidth]{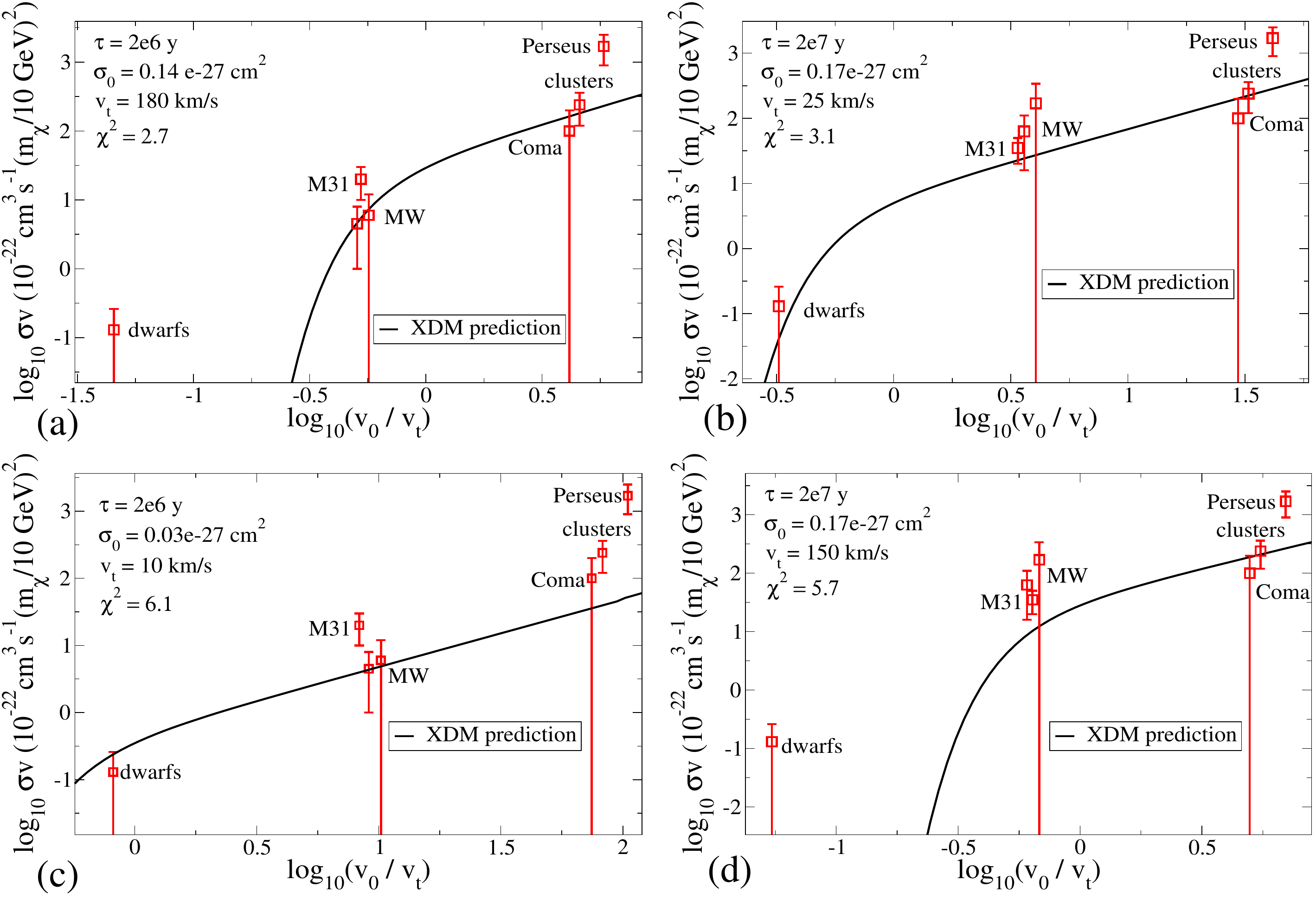}}
\caption{Data points: experimentally motivated values of the cross section
(times $(10{\rm\ GeV}/m_\chi)^2)$ versus $v_0/v_t$ (proportional
to DM velocity dispersion, $\sigma_v = \sqrt{3/2}\,v_0$).  Curve is theoretical prediction of excited dark
matter framework.  Left (right) column is for excited state lifetime
$\tau = 2\times 10^6\,$ y ( $2\times 10^7\,$ y). 
Topmost graphs are for best-fit values of $\sigma_0$ and
$v_t$ in fig.\ \ref{cont-panel}).  Bottom graphs are for the
lower or higher values of $v_t=10,\,150$ km/s relative to best fits
(also acceptable fits), the first being marginally
compatible with limit from dwarf spheroidals.}
\label{fit-panel}
\end{figure*}

Inserting the approximate correction factors into the 
(numerically evaluated) integrals of $\rho^2/x^2$
for the two galaxies, we find no significant change in the predicted
flux from M31 except for $\sigma=0.1\,r_s$, where it decreases by
only a factor of 1.9.  On the other hand,  for $\sigma = 
(10^{-2},\,10^{-1})\, r_s$
there is a respective factor of $(18,\,300)$ reduction in the
flux from
the MW for the FOV of ref.\ \cite{Riemer-Sorensen:2014yda}, requiring a corresponding 
increase in the upper limit for the cross
section, $(0.15-1.1)\to (3-20)$ or $(40-330)$ as indicated in
column 5 of table
\ref{tab1}.  For the FOV of ref.\ \cite{Boyarsky:2014ska}, the
reduction is a factor of $(12,\,200)$ respectively,
leading to the replacement
$(0.1-0.7)\to (2-12)$ or $(20-130)$ in the allowed values of
$\langle\sigma v\rangle$. 
For streaming lengths $\sigma$ closer to $0.01\,r_s$ 
the target values of $\langle\sigma
v\rangle$ become comparable for M31 and MW, as would be expected for
two such similar galaxies, while the range of upper limits is also
compatible with these detections.  On the other hand, fluxes from clusters should be
unchanged due to the much smaller value of $\sigma /r_s$ in those 
much larger systems.  Similarly, galaxy limits of ref.\ 
\cite{Anderson:2014tza} and \cite{Malyshev:2014xqa} are unchanged since the FOVs cover a much
greater fraction of $r_s$ than for the MW observation.  Streaming
lengths $\sigma\lesssim 10^{-3}\,r_s$ are too short to make any difference
for the MW fluxes.

We do not attempt a more quantitative fit of the smearing length here,
in light of the large uncertainties in the desired cross section
values.  The approximate lifetime needed for the excited state is 
$\tau \sim \sigma/\sigma_v \cong (0.2,\,2)\,$kpc/(100
km/s) $\cong 2\times 10^6\,$y or $2\times 10^7\,$y respectively,
for $\sigma = (0.01,\,0.1)\,r_s$.  It will be seen presently that
the shorter lifetime gives a somewhat better fit to the data.

To fit the resulting cross sections together with those from other systems, we need to
know the velocity dispersions of the two galaxies.  For the MW,
data exists only down to $r\cong 10$ kpc, where estimates range from
$\sigma_v = 105$ to 130 km/s \cite{Dehnen:2006cm,Brown:2009nh}.  We have
adopted an average value 118 km/s.
For M31, measurements exist for smaller radii $\sim 0.1\,$kpc
\cite{merrett}.  We reproduce these measurements in fig.\ 
\ref{M31-fig}, along with our fit to the analytic form
$\sigma_v = (57.9 + 108.3\,e^{-5.05\,\theta})\,$km/s, where $\theta$
is in degrees.  Translating this into a radial dependence, we compute
the average value $\langle\sigma_v\rangle = [\int d^{\,3}x\,
\rho^2\sigma_v/x^2]/[\int d^{\,3}x\, \rho^2/x^2]$ $= 116\,$km/s,
where the angular part of the integrals is over the $0.22^\circ$ FOV.
This procedure is only meaningful for estimating the effect on the
cross section through eq.\ (\ref{gamma_eq}) if $v_t \lesssim 
\sqrt{2/3}\langle\sigma_v\rangle$; otherwise we should compute the
average value of $\gamma$ in the integral rather than just $\sigma_v$
(for small $v_t$ the two are proportional).   We have carried this out
for the 0.22$^\circ$ FOV and the result is shown in fig.\ 
\ref{M31-gamma}.

\subsubsection{Noncuspy halo profiles}
\label{noncuspy}

Table \ref{tab1} shows that the cored Burkert profile goes
too far in reducing the MW signal relative to that of M31.  A dark
matter profile somewhere between Burkert and NFW is needed to give
maximum overlap between the desired ranges of cross sections for the 
two galaxies, assuming that the excited state decays promptly.  An 
example that can interpolate between the two is the Einasto profile,
\be
	\rho_E = \rho_s \exp\left(-{2\over\alpha}\left[ 
	\left(r\over r_s\right)^\alpha
	- 1 \right]\right)
\ee
A set of standard values often used, compatible with predictions
from DM-only $N$-body simulations \cite{Navarro:2008kc}, 
is $\alpha=0.17$, $r_s=20\,$kpc.  The
concentration of DM near the center depends strongly upon $\alpha$,
and we find that a large value $\alpha=0.30$ (holding $r_s$ fixed
at 20 kpc) is required to increase the cross section for MW
observations of the X-ray line by a factor of 36 while increasing that
of M31 by only a factor of 2, which may be sufficient to reconcile their 
values within the errors. With larger $r_s=30\,$kpc, smaller $\alpha=0.25$ 
can achieve a similar effect.

The question of whether the MW halo is cuspy or cored is still debated
in the literature, with some indications that including the effects of
baryons leads to cuspier halo 
\cite{Tissera:2009cm},\cite{Schaller:2014uwa},
while others argue that cored profiles are observationally preferred
\cite{Nesti:2013uwa}.  We consider the noncuspy possibility as one
option for understanding the strength of the 3.5 keV line from the
galactic center, and will use this freedom to fit it away using the
unknown halo profile in part of the analysis that follows.

\begin{figure}[t]
\centerline{\includegraphics[width=1.25\columnwidth]{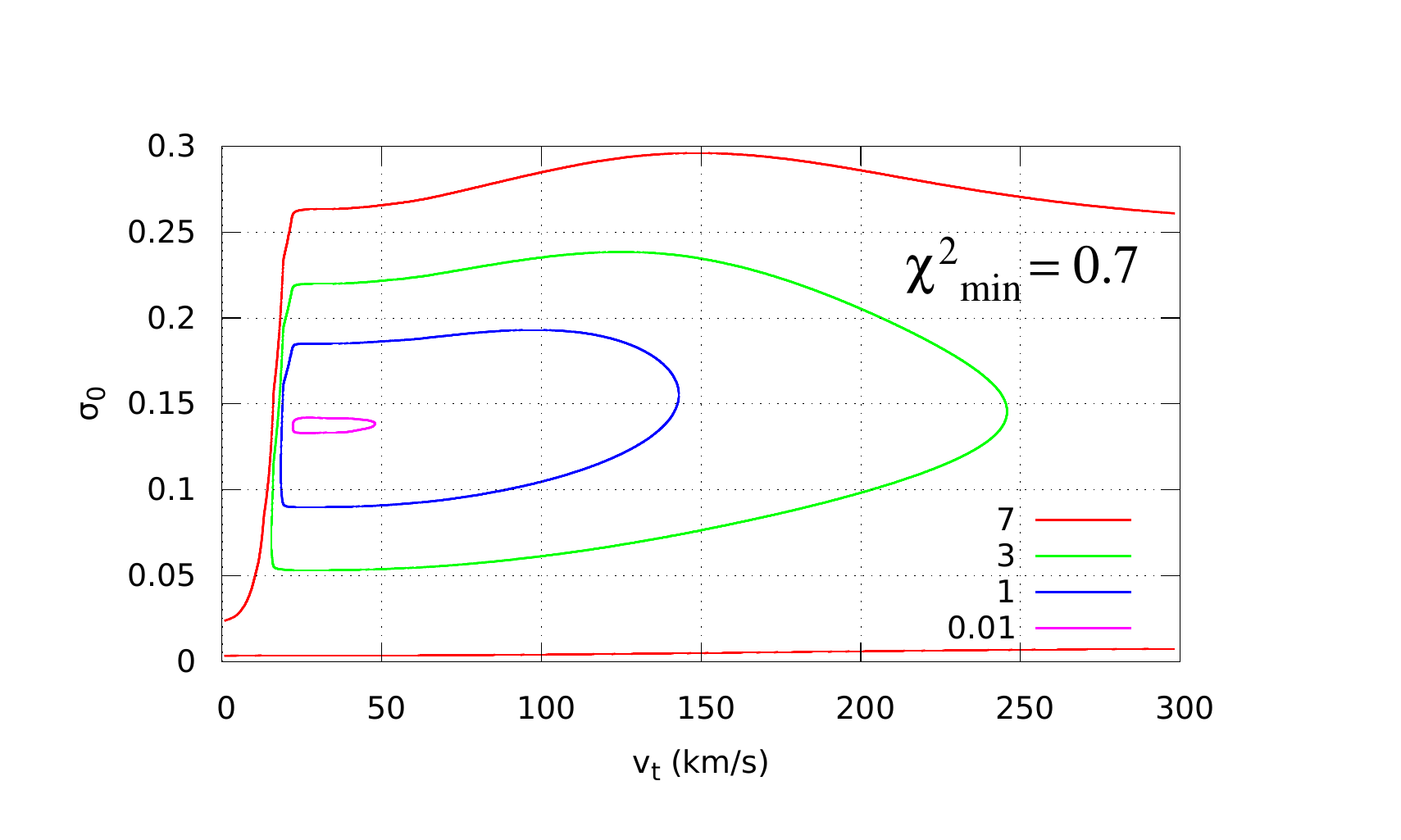}}
\caption{Similar to fig.\ \ref{cont-panel} but for promptly decaying 
excited states, with MW contributions assumed to be fitted by
adjusting the DM halo profile.}
\label{cont-fast}
\end{figure}

\begin{figure*}[t]
\centerline{\includegraphics[width=2\columnwidth]{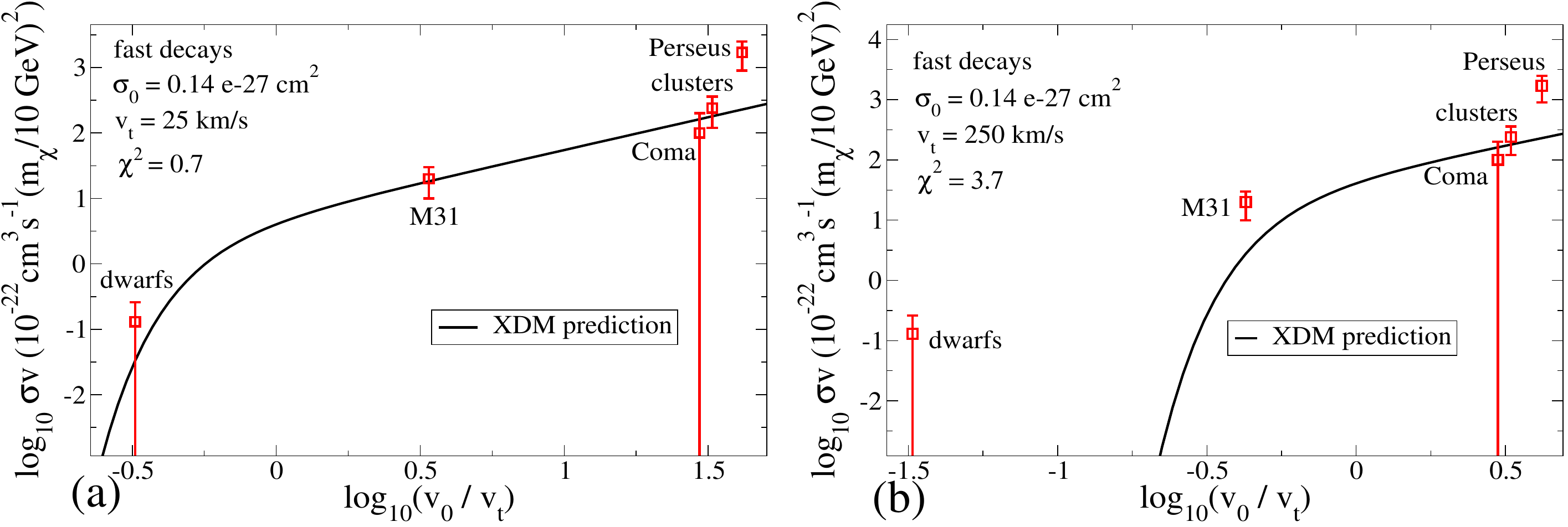}}
\caption{Similar to fig.\ \ref{fit-panel} but for promptly decaying 
excited states, with MW contributions assumed to be fitted by
adjusting the DM halo profile.}
\label{fit-panel-fast}
\end{figure*}

\begin{figure}[t]
\centerline{\includegraphics[width=\columnwidth]{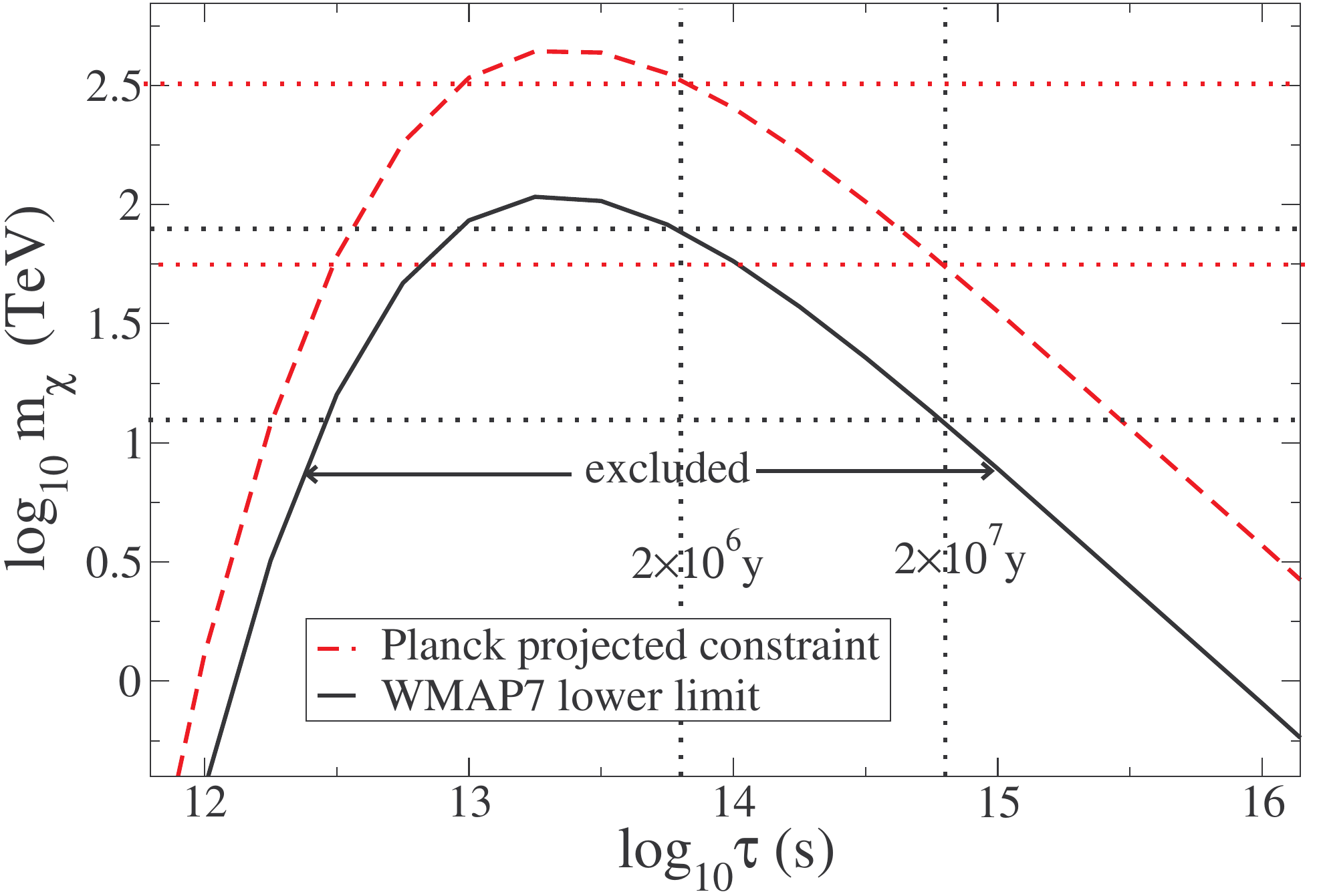}}
\caption{CMB Lower limit on DM mass $m_\chi$ versus lifetime of excited
state, assuming relative abundance $f_x=1/2$ of the excited state.}
\label{cmb-limit}
\end{figure}

\subsection{XDM parameter best fits}

We proceed to search for the preferred values of the XDM model
parameters $v_t$, the threshold velocity, and the cross section
normalization $\sigma_0$, defined in eqs.\ (\ref{vteq},\ref{xsecteq}),
using the cross section values in columns 4 and 5 of table \ref{tab1}.
We consider two cases: (1) the 
excited state has lifetime $\sim 10^6-10^7\,$y, and the MW has an NFW
profile; (2) the excited state undergoes fast decays,
and the MW halo is assumed to have the right shape for resolving the
tension between MW and M31 observations of the X-ray line.   
In the first case we include MW data in the fit while
in the second we omit them.

\subsubsection{Long-lived excited states}

Starting with the intermediate lifetime scenario, 
recall that the values in column 4 are unchanged for sources that have no
entry in column 5, and the upper (lower) entries of the latter apply
for the excited state lifetime of $2\times 10^{6}(10^7)\,$y; we consider both
cases here.  The predicted value of the cross section is given by
eqs.\ (\ref{xsecteq},\ref{gamma}) with $v_0 = \sqrt{2/3}\,\sigma_v$
and $\sigma_v$ given in column 7 of table \ref{tab1}, except for M31
where we use the more quantitative averaging of $\gamma$ over the FOV
as described in section \ref{M31sect} and fig.\ \ref{M31-gamma}.\footnote{In order
to meaningfully display M31 data in figs.\ \ref{fit-panel} and
\ref{fit-panel-fast}, we have adjusted its value of $v_0/v_t$ on the plots so
that the deviation from the simple predicted curve shown there
matches the deviation from the actual prediction using fig.\ \ref{M31-gamma}.}  
As previously discussed, we do not include the measurements of the 
off-center Perseus cluster flux in our fits since they are not compatible with
the model (nor with decaying DM models, if the dwarf spheroidal
constraints are believed).

Defining a $\chi^2$ statistic using the cross section ranges shown in
table \ref{tab1} to estimate the central values and the uncertainties, we find preferred
regions in the $v_t$-$\sigma_0$ plane as shown in fig.\
\ref{cont-panel}.  The uncertainties are presently too large to justify
a formal statistical analysis, so instead of showing the usual
confidence intervals, steps in $\delta\chi^2$ of $\sim 1,\,2,\,4$ are chosen for the 
contours above the minimum value.  The best-fit values of the threshold
velocity are 180 (25) km/s for the $10^{6}(10^7)\,$y lifetimes,
respectively, not including the Perseus cluster in the fit; including
it shifts these values to 186 (25) km/s.  
Using (\ref{vteq}), these correspond to the range of 
dark matter masses $m_\chi \sim 80$ GeV to $4$ TeV, with $m_\chi$ going as
$v_t^{-2}$. 
The minimum value of
$\chi^2$ increases by approximately 3 when including Perseus on-center
data in the
fit, but the best-fit values do not shift significantly.
The comparisons of the
predictions to the data, in terms of $\langle\sigma v\rangle$ versus
$v_0/v_t$, are shown in figs.\ \ref{fit-panel}(a,b) for these two
lifetimes.  The predicted signal is far below the sensitivity of 
spheroidal dwarf searches at low $v_0$ for $\tau=2\times 10^6\,$y, but
saturating the dwarf constraint for $\tau=2\times 10^7\,$y.

However the minimum of $\chi^2$ is rather shallow and allows for 
larger or smaller values of $v_t$ to still give an acceptable fit.  
Thus the correlation of smaller $\tau$ with larger $v_t$ is not
mandatory, and we display examples with the opposite behavior yet
still providing acceptable fits in  figs.\ \ref{fit-panel}(c,d),
with $v_t$ as low as $10$ km/s, corresponding to dark matter mass
$m_\chi = 25$ TeV.
Thus in either model  it would be possible to start to detect the
X-ray line in dwarfs given longer exposures, or for the signal from
dwarfs to be far too weak for detection. The range of allowed DM
masses remains as previously estimated.

\subsubsection{Fast-decaying excited states} If we assume that the MW halo has the
right shape for consistency of the X-ray limits or detections from the galactic
center, as we have already argued is plausible, then the MW data can be omitted
from our  fits, and we obtain the $\chi^2$ contours shown in fig.\ 
\ref{cont-fast}.  The best fit parameters are similar to those in the
previously considered cases of longer excited state lifetimes, and also like those
cases, acceptable fits can be found at either high or low  threshold velocities
where dwarf spheroidals could be respectively far from or close to providing an
observable strength of the X-ray line.  This is illustrated in fig.\
\ref{fit-panel-fast}, showing the data versus the prediction for cross section
versus DM velocity.  Allowing for $v_t$ in the range $25-300$ km/s, we find the
range of dark matter masses 40 GeV to 30 TeV, with large masses corresponding to
low threshold velocities.

\begin{figure*}[t]
\centerline{\includegraphics[width=0.9\columnwidth]{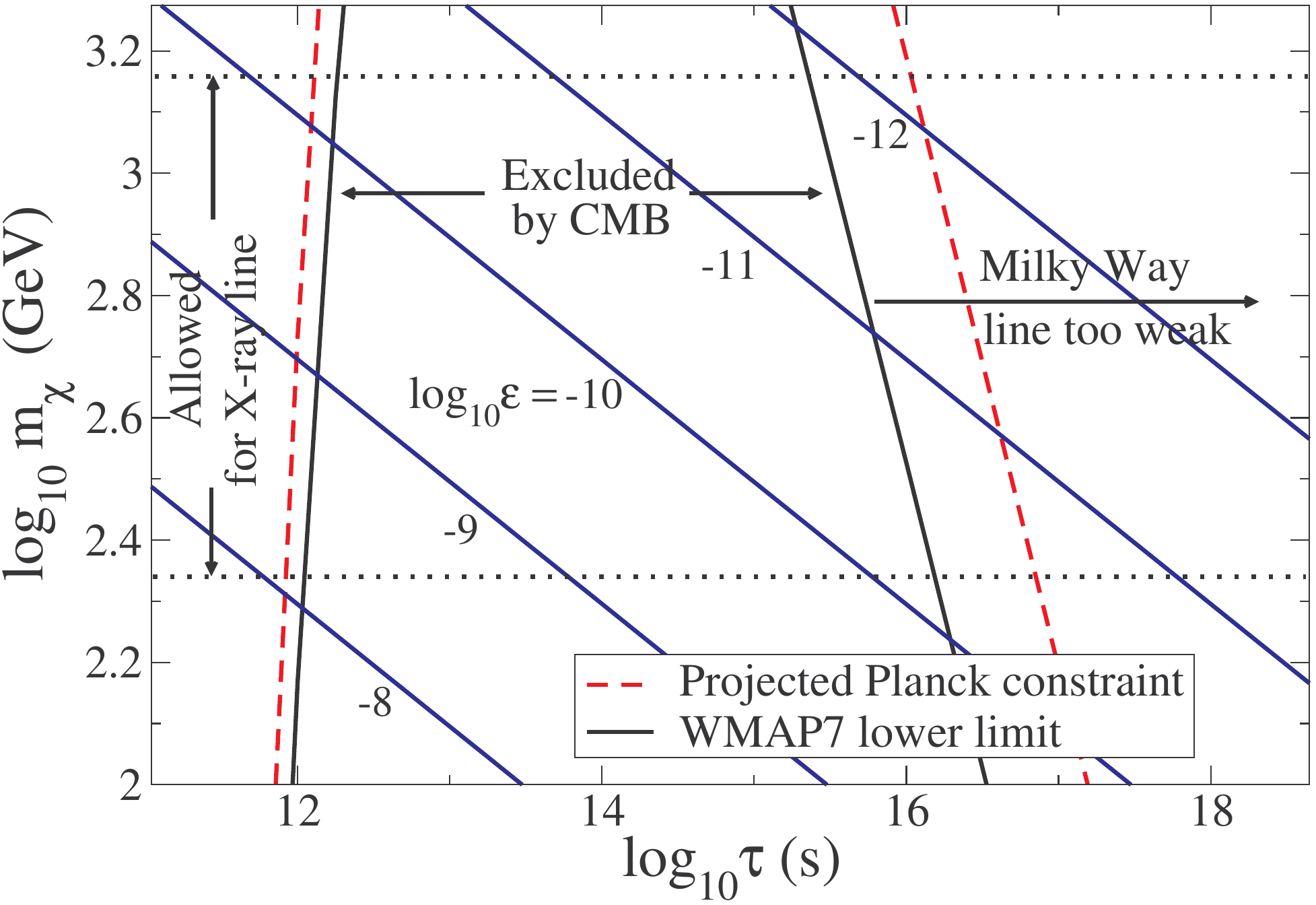}
\includegraphics[width=1.01\columnwidth]{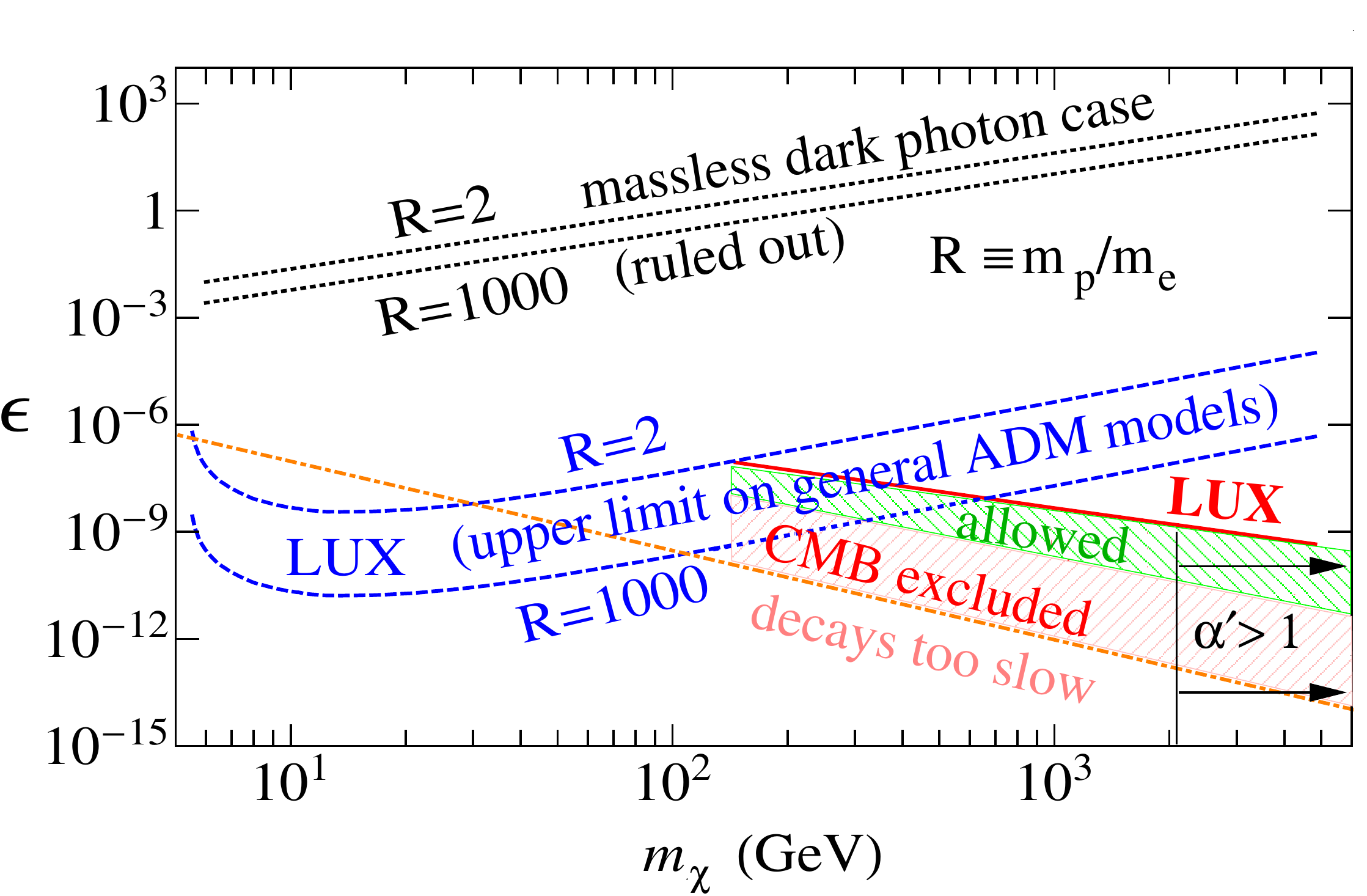}}
\caption{Left: similar to fig.\ \ref{cmb-limit}, with diagonal lines 
showing contours of constant gauge kinetic mixing parameter $\epsilon$
in the atomic DM model.  Right: impact of CMB constraint on region 
allowed by direct detection in $m_\chi$-$\epsilon$ plane.}
\label{cmb-atomic}
\end{figure*}

\subsection{CMB constraints}

There is one further experimental constraint on XDM models that must
be considered when the excited state lifetime is as long as those we
have investigated.  Such decays in the early universe inject electromagnetic
energy into the thermal plasma at a time when it can have an impact on
the temperature fluctuations of the cosmic microwave background by
delaying recombination, changing the optical depth, or the redshift of
reionization \cite{Chen:2003gz}-\cite{Diamanti:2013bia}.  If the
lifetime is sufficiently short or long, the decays either complete
before the relevant epoch or are too slow to have an appreciable effect.  But
for the lifetimes we have singled out as being interesting for
reconciling X-ray observations of the MW galactic center with other
measurements of the 3.5 keV line, decays producing photons can have a
significant impact on the cosmic microwave background (CMB).

The constraints on decays are typically expressed as an upper limit on
the fraction of the total DM energy density that can released into 
photons (or other ionizing radiation).  This fraction is given by
\be
	f = f_{x} {\Delta E\over m_\chi}
\ee
where $f_x$ is the abundance of the excited state relative to the
total DM abundance, and $\Delta E = 3.5$ keV for the current
application.  An upper limit on $f$ thus translates into a lower limit
on the DM mass $m_\chi$.  We have estimated this limit for the case
$f_x =1/2$ as a function of the excited state lifetime in fig.\
\ref{cmb-limit}. 
The current hard constraint is based upon WMAP7 data, while projected 
constraints using Planck data await the release of Planck polarization
results.  We have determined these constraints using
the methods described in ref.\ \cite{Cline:2013fm}.  The latter are
limited to injection energies no less than $\sim 5$ GeV, while
we are interested in the case of 3.5 keV.  We have rescaled the
constraints for the case of DM decaying into two 5 GeV photons
by a factor of $8.3/2 = 4.2$ greater sensitivity, by reading off from
fig.\ 6 of ref.\ \cite{Slatyer:2012yq} the relative constraint on the lifetime as
a function of deposited energy.

From fig.\ \ref{cmb-limit} we see that large DM masses are required,
$m_\chi>5$ TeV, for the least constraining case of $\tau=2\times
10^7\,$y, using the WMAP7 data, while at $\tau=2\times
10^6\,$y the WMAP7 limit is $m_\chi > 32$ TeV, both being
somewhat in tension
with our low-$v_t$ fits to the X-ray line strength, notably the limit
from nonobservation in dwarf spheroidals.  If the projected Planck
limits are validated, then the scenario is ruled out, 
with the limit $m_\chi > 55$ TeV in the longer lifetime case
being incompatible with the dwarf constraints.

One must keep in mind however that these constraints would be relaxed
if there is a mechanism for depleting the excited state relative
abundance $f_x$ before the epoch of decays in the early universe.
For example if the kinetic temperature of the DM becomes lower than
that of the visible sector at early times, the relaxation process 
$\chi' \chi'\to \chi\chi$ can effectively deplete the excited state,
before the redshifts relevant for the CMB, due to the relatively high
DM density.  This requires that inelastic scattering $\chi+X\to
\chi'+X$ on standard model particles $X$ go out of equilibrium at
early times to prevent the repopulation of the excited state.
After structure formation, following $\chi \chi\to \chi'\chi'$ in 
galaxies, the relaxation process can be slower than decays because of 
the smaller excited state density relative to that in the early universe.

\section{Models}
\label{models}
We briefly consider the implications of our constraints for some
specific models of XDM that have been previously proposed for the
3.5 keV line.  In ref.\ \cite{Cline:2014eaa}, the hyperfine transition
in a model of atomic DM with kinetic mixing of the dark photon to the
normal photon was suggested, in which the cross section for
$\chi\chi\to\chi'\chi'$ was estimated to be $100\, a_0^2$, where $a_0$
is the Bohr radius of the dark atom.  We identify this with the
parameter $\sigma_0 \cong 0.2\times 10^{-27}$cm$^2(m_\chi/10{\rm\
GeV})^2$ determined from our fits in the previous section.  The
mass splitting $\Delta E= 3.5$ keV is predicted to be $(8/3)\alpha'^4
m_\chi/f(R)^2$, where $\alpha'$ is the dark gauge coupling, $R$ is the ratio of the dark proton to dark
electron mass, and $f = R + 2 + 1/R\ge 4$.  Eliminating $\alpha'$ from
these relations, one can solve for the dark atom mass 
$m_\chi = 90 (f/4)^{2/7}$ GeV, smaller by a factor of $0.65$ than the
estimate made in \cite{Cline:2014eaa}.  For perturbativity in
$\alpha'$, one requires $R < 5\times 10^4$ hence $m_\chi < 1.4$ TeV,
while efficient recombination in the dark sector requires $R\gtrsim
100$ hence $m_\chi > 225$ GeV.  
The entire range of allowed values for the DM mass in this model is 
therefore consistent
with the current data for the line (excluding the Perseus cluster as
we have discussed).  

CMB constraints on the atomic XDM model were not considered in 
ref.\ \cite{Cline:2014eaa}.  The lifetime of the excited state is
predicted to be $\tau = 3 m_\chi^2 / (\alpha\epsilon^2 f^2 \Delta
E^3)$ where $\epsilon$ is the gauge kinetic mixing parameter.  
We illustrate the CMB constraints over the relevant mass range 
by plotting contours of constant $\epsilon$ over the previously
shown CMB upper limit on $m_\chi$ as a function of $\tau$ in 
fig.\ \ref{cmb-atomic}(left).  The regions between the dotted horizontal
lines and to the left of the CMB curve are allowed.
The intermediate lifetime cases $\tau\sim 2\times 10^6-2\times 10^7$y
($10^{13}-10^{15}$s) are excluded for the atomic XDM model, and longer
lifetimes are disfavored by the claimed Milky Way observations since
they would dilute the signal from the GC too much.  We must rely upon
a noncuspy halo profile in this case, as discussed in section 
\ref{noncuspy}.  The CMB constraint significantly reduces the region
of parameter space in the $m_\chi$-$\epsilon$ plane that is allowed
by direct detection, as shown in fig.\ \ref{cmb-atomic}(right).
The model can eventually be ruled out or discovered by improvements in sensitivity
of direct DM searches.

A second class of realizations of XDM was provided in ref.\ 
\cite{Cline:2014kaa}, where the dark sector has a broken
SU(2) gauge symmetry with nonabelian kinetic mixing between
one of the dark gauge boson components and the photon.  In these
models, the kinetic mixing parameter $\epsilon$ must be sufficiently 
large to get the observed X-ray line strength, leading to 
stronger direct detection constraints on 
$\epsilon$, and the necessity to demand $m_\chi\lesssim$ a few GeV
to evade these constraints.  Through eq.\ (\ref{vteq}) this 
corresponds to large threshold velocities $v_t\gtrsim 1100\,$km/s
that are strongly disfavored by our fits since they would suppress
signals from any sources except for galaxy clusters.  These models thus
seem to be in conflict with the current data.  It should be kept in
mind however our assumption that all the significant velocity
dependence in the cross section is due to the phase space.  Models
with very light mediators, hence an additional source of velocity
dependence, require special treatment that is beyond the scope of
the present work.

\section{Conclusions}

In this paper we have reassessed the observational claims for and
against the 3.5 keV X-ray line, with emphasis on the possibility that
excited dark matter models can overcome the discrepancies that
decaying DM models seem to exhibit. Although the flux from the 
Perseus cluster is somewhat too high compared to the other sources,
especially at large angles from the cluster center, the error bars for
the on-center data are sufficiently large that it is possible to
obtain a reasonable fit to these data, combined with stacked galaxy
clusters, galaxies, M31, the galactic center, and dwarf spheroidals,
within the XDM framework.  
While XDM does introduce a new parameter to DM models (the threshold 
velocity $v_t$), it is noteworthy that the data roughly follows the expected
dependence on the DM velocity dispersion and removes the contradictory
observations between low- and high-dispersion objects.
In order to resolve a slight discrepancy between M31 and MW observations, 
it is possible to introduce another parameter, the lifetime of the
excited state decay; however, it is also possible that a slightly cored
DM halo for the MW resolves this discrepancy.  In fact, this second 
option is preferred in the concrete XDM model we considered here and is well
within the limits of our knowledge about the DM distribution in the MW.

Nonobservation of the line in dwarf galaxies is consistent with
expectations that objects with low DM velocity dispersion should give
a smaller signal from XDM.  However the current data allow a large
range of dark matter masses, $m_\chi \sim 40$ GeV $-$ 25 TeV.  At the
heavy extreme, the threshold velocity for producing the excited state
can be sufficiently low so that dwarf spheroidals may be close to
exhibiting a positive signal for the line, given longer exposures 
than those used so far to obtain upper limits on the flux.  At the
lighter end, these systems will always be orders of magnitude below
the required sensitivity, whereas it is galaxies like M31 and the MW
that are close to the kinematic threshold for producing the excited
states.  Hopefully recent proposals to observe the line more carefully
will lead to clarification of the experimental situation in the near
future, and enable us to better constrain this class of models.

\bigskip
\bigskip
{\bf Acknowledgments.}  We thank A.\ Boyarsky, E.\ Bulbul,
G.\ Holder, D.\ Malyshev, 
G.\ Moore, O.\ Ruchayskiy, P.\ Scott,
R.\ Riemer-S\o rensen, T.\ Slatyer, O.\ Urban and A.\ Vincent for helpful 
discussions or correspondence.  We thank J.\ Conlon for pointing 
out an error in the first version of this paper.
Our work is supported by the Natural
Sciences and Engineering Research Council (NSERC) of Canada.

\appendix
\section{Profile integrals}  Here we give expressions for various integrals over the
DM density.  The functions $h_n$ from eq.\ (\ref{gamma_eff}) are given
by
\bea
	h_1 &=& \ln(1+c) - {c\over 1+c}\nonumber\\
	h_2 &=& \sfrac13\left(1-(1+c)^{-3}\right)
\eea
for NFW profiles, while for the Einasto profile $\rho = \rho_s
e^{-(2/\alpha)((r/r_s)^\alpha-1)}$,
\be
	h_n = {e^{2^n/\alpha}\over\alpha}\left(
8^{-{n\over\alpha}}\,\alpha^{3\over\alpha}\,\Gamma_\nu\left(
\sfrac{3}{\alpha}\right)
	-c^3\, E_{1-{3\over\alpha}}\left({2^n\,c^\alpha\over\alpha}\right)
	\right) 
\ee

The integrals of $\rho/x^2$ and $\rho^2/x^2$ over a field of
view can be
performed analytically for NFW profiles using two approximations.
First one writes 
\be
	\int{d^{\,3}x\over x^2} \rho^n \cong 2\pi\int_{\cos\theta_0}^1
	\!\!\!\! d y
	\int_{-\infty}^\infty \!\!\!dz\, \rho^n(\sqrt{z^2 + d^2 - 2 z
	y d })
\ee
where $\theta_0$ is the angular size of the observed region
and $y=\cos\theta$.
A small error is made by including the region behind the
observer in the LOS integral; this makes the $z$ integral analytically
tractable after shifting $z\to z+yd$ and completing the square so
that $z^2 + d^2 - 2 z 	y d \to z^2 + d^2(1-y^2) = z^2 +
d^2\sin^2\theta$.  Then one makes the small-angle approximation $y =
1-\theta^2/2$ (or $\sin\theta=\theta$) and integrates with respect 
to $\theta$.  The result is
\be
\int{d^{\,3}x\over 4\pi\,x^2}\rho^n \cong 
	{\rho_s^n r_s^3\over d^2}\, f_n(a)
\label{fneq}
\ee
where $a = d\theta_0/r_s$ and the dimensionless functions are given
by
\bea
	f_1 &=&  {\rm Re}\left[ \ln(a/2) + {\ln\left((1 +
\sqrt{1-a^2})/a\right) \over
	(1-a^2)^{1/2} }	\right]\nonumber\\
	f_2 &=&
{\rm Re}\left[\frac{4-a^2}{6\left(a^2-1\right)^2}
+{\pi a\over 2} -\frac23 \right.\nonumber\\
&-&\left.\frac{\left(2 a^4-5a^2+4\right) 
a^2 \tanh^{-1}\left(\sqrt{1-a^2}\right)}{2\left(1-a^2\right)^{5/2}}
\right]
\label{fieqs}
\eea
The above expressions are manifestly real if $a<1$, and their analytic
continuation to $a>1$ is correctly given by taking the real parts,
which we find simpler than specifying the real analytic continuations
explicitly.

The above treatment assumes that the field of view is centered on the
object of interest.  If the FOV is off-axis by an angle $\theta$ which is 
much larger than the opening angle of the FOV, then the above treatment
is modified; instead of integrating over $\theta$ (and the azimuthal
angle), one simply multiplies by the solid angle $\delta\Omega$ 
of the FOV.  In this case we obtain 
\be
\int{d^{\,3}x\over 4\pi\,x^2}\rho^n \cong 
	{\rho_s^n\, r_s}\,{\delta\Omega\over 4\pi}\, g_n(a)
\ee
where $a=d\sin\theta/r_s$ and 
\bea 
	g_1 &=& {\ln\left((1 + \sqrt{1-a^2})/a\right) \over
	(1-a^2)^{3/2} } - (1-a^2)^{-1/2}\nonumber\\
	g_2 &=& {-6 a^7+23 a^5-43 a^3+ 26 a\over
	6 a
   \left(1-a^2\right)^4} + {\pi\over 2a}\nonumber\\
&+& {\left(2
   a^6-7 a^4+8 a^2-8\right)\tanh ^{-1}\left(\sqrt{1-a^2}\right)
\over 2(1-a^2)^{7/2}}\nonumber
\eea

\end{document}